\documentclass[11pt,a4paper]{article}

\usepackage{amsmath}
\usepackage{amssymb}
\usepackage{amsfonts}
\usepackage{amscd}
\usepackage{amsthm}
\usepackage{xypic}
\usepackage{enumerate}
\usepackage{fancyhdr}
\usepackage{color}
\usepackage{graphicx}
\usepackage{subfig}
\usepackage{tikz}
\usepackage{pgf}
\usepackage{hyperref}
\usepackage{tabularx,colortbl}
\usepackage{biograph}
\usetikzlibrary{snakes}

\newcommand{\R}{\mathbb{R}}

\newcommand{\oF}{\overline{F}}
\newcommand{\oS}{\overline{S}}
\newcommand{\oE}{\overline{E}}

\DeclareMathOperator*{\im}{im}
\DeclareMathOperator*{\Con}{Con}

\numberwithin{equation}{section}
\newtheorem{lemma}[equation]{Lemma}
\newtheorem{theorem}[equation]{Theorem}
\newtheorem{proposition}[equation]{Proposition}
\newtheorem{corollary}[equation]{Corollary}

\theoremstyle{definition}

\newtheorem{remark}[equation]{Remark}

\begin{document}

\title{An Algebraic Approach to Signaling Cascades with $n$ Layers}

\author{Elisenda Feliu$^{1,2}$, Michael Knudsen$^2$, Lars N. Andersen$^2$, Carsten Wiuf$^2$}

\maketitle

\footnotetext[1]{Facultat de Matem\`atiques, Universitat de Barcelona, Gran Via 585, 08007 Barcelona, Spain}
\footnotetext[2]{Bioinformatics Research Centre, Aarhus University, C. F. M{\o}llers Alle 8, Building 1110, DK-8000 Aarhus C, Denmark}

\begin{abstract}
Posttranslational modification of proteins is key in   transmission of  signals in cells. Many signaling pathways contain several layers of modification cycles that mediate and change the signal through the pathway. Here, we study a simple signaling cascade consisting of $n$ layers of modification cycles, such that the modified protein of one layer acts as modifier in the next layer. Assuming  mass-action kinetics and taking the formation of intermediate complexes into account, we  show that the steady states are solutions to a polynomial  in one variable, and in fact that there is exactly one steady state for any given total amounts of substrates and enzymes. 

We demonstrate that many steady state concentrations are related through rational functions, which can be found recursively. 
  For example, stimulus-response curves arise as inverse functions to explicit rational functions. We show that the stimulus-response curves of the modified substrates are shifted to the left as we move down the cascade. Further, our approach allows us to study enzyme competition, sequestration  and  how the steady state changes in response to changes in the total amount of substrates.

 Our approach is essentially algebraic and follows recent trends in the study of posttranslational modification systems.
 
\end{abstract}

\section{Introduction}

Posttranslational modification of proteins is one of the principal mechanisms by which signals are transmitted in living cells.
In particular, covalent modification by phosphorylation is wide-spread and provides one of the basic modes by which a signal is mediated through a pathway. 
Classical signaling pathways typically contain  a cascade of post-translational modification cycles (also called futile cycles) where the activated protein in one layer acts as the modifier enzyme in the next layer. In order to understand the biological importance of this sophisticated signaling mechanism and to predict the behavior of the system, a number of theoretical studies have focused on determining the system's dynamics and steady states. In the most discussed case, the MAPK cascade, the presence of bistability and oscillatory behavior has been revealed for different choices of system parameters \cite{Ferrell-bistability,Qiao-bistability}. 

The basic features of   general signaling cascades can be elucidated from the study of signaling cascades with $n$ layers and  one cycle of post-translational modification at each layer. These cascades are part of many pathways  (examples include the cAMP pathway \cite[page 342]{cell},  quorum sensing pathways \cite{quorum} and the intrinsic cascade in the blood coagulation pathway \cite{macfarlane}) and  have been extensively studied from different mathematical points of view:  $n=1$, e.g.~\cite{Bluthgen-sequestration,salazarN1}, $n=1,2$, e.g.~\cite{Gold-Kosh-81,Gold-Kosh-84,bluthgen-switch}, and arbitrary $n$, e.g.~\cite{Sontag-cascade,khodolenko-97,Vondriska-cascade,Ventura-Hidden}.
Here, we use the classical model of Michaelis and Menten  in which the enzyme-substrate complex is formed reversibly and  the dissociation of the complex into a modified substrate and a free enzyme is assumed to be irreversible. 
 Further, the modifier, which in the case of phosphorylation typically would be ATP, is  assumed to be in constant concentration and hence not included explicitly in the model. In this we follow much previous work, e.g.~\cite{Gold-Kosh-81,G-PNAS,Ventura-Hidden}.

In order to simplify the mathematics of the Michaelis-Menten model, additional requirements are often  imposed; e.g. the formation of intermediate complexes is ignored under the quasi steady state assumption \cite{Sontag-cascade,Heinrich-kinase,Vondriska-cascade} or the total amount of substrate is  assumed to be much higher than those of the (de)modification enzymes \cite{Ventura-Hidden}.  In the first case, this has the consequence that some biological mechanisms, like sequestration, might be  overlooked, while the second case is not, in general, considered biologically valid \cite{Bluthgen-sequestration}. Additionally, conclusions are often drawn from parameter sampling and are not based on a rigorous mathematical analysis of the model \cite{Racz-amplification}.

We focus on understanding the steady state behavior of  a signaling cascade with $n$ layers and one  cycle of post-translational modification at each layer. Using mass-action kinetics and without imposing further assumptions on the Michaelis-Menten model, we show that the steady states arise as the solutions to  one polynomial in one variable. This allows us to conclude that these systems have exactly one steady state for fixed total amounts of substrates and enzymes. This is achieved using an induction argument that reflects the modularity of the cascades.

 Our approach makes it possible to study aspects of the system in detail without relying on simulation. For example, the stimulus-response curve arises as the inverse function of an explicit rational function (i.e. a quotient of polynomials in one variable). Several other rational functions are found that describe the species concentrations at steady state; these functions provide a clear picture of how the steady state re-accommodates  to variations in initial concentrations and provide insight into sequestration. In addition, our approach enables us to explore the parameter space in more detail than previously and to divide the space into regions that exhibit different steady state behaviors.

The simplicity exposed by the algebraic approach is the key component in this work and follows the line of  recent works \cite{G-PNAS,TG-rational,TG-Nature}, as well as some older approaches \cite{bardsley-sigmoid}. All rational functions can be explicitly computed. In addition, their inverses can be obtained using any program that allows symbolic algebraic computations like Mathematica\texttrademark{}.

The article is divided into two main sections. In the first the mathematical theory is developed. We show that the steady states can be described as the solutions to a polynomial in one variable only and that there is  only one steady state for any initial total amounts of substrates and enzymes. Subsequently, we  derive rational functions that relate the concentrations of chemical species at steady state. In the second section, we use these results to  discuss various topics of biological interest such as enzyme competition, stimulus-response curves, and  how steady state concentrations vary in response to changes in the total amount of substrates. We end with some concluding remarks. Proofs are given in the appendix.

\bigskip
{\small
\emph{Acknowledgements.}
This work was done while EF was visiting Aarhus University in Spring 2010. It was completed while EF and CW were visiting Leipzig University in July 2010. EF wishes to thank the members of the Bioinformatics Research Centre in Aarhus for their hospitality. 
EF is supported by the Spanish Ministerio de Ciencia e Innovaci\'on by project MTM2009-14163-C02-01 and a travel grant from the Universitat de Barcelona. MK, LNA and CW are supported by the Danish Research Council and CW by the Humboldt Foundation.  All authors are supported by a travel grant from the Carlsberg Foundation, Denmark.  Joaquim Puig is thanked for useful discussion.}

\section{Mathematical results and characterization}
\label{math_sec}

\subsection{Preliminaries}
By a \emph{rational function} $f(x)$ over a field $K$ we understand a function that is given as a quotient of polynomials $f(x)=p(x)/q(x)$ with coefficients in $K$. 
Assuming that it is in reduced form, that is, there are no common zeros between $p(x)$ and $q(x)$, then the only discontinuities of the function are given by the zeros of the denominator. Note that two reduced forms $p_1(x)/q_1(x)$ and $p_2(x)/q_2(x)$ of the same rational function satisfy $p_1(x)=\lambda p_2(x)$ and $q_1(x)=\lambda q_2(x)$ for some $\lambda\in K$.
If not otherwise stated, we consider rational functions over $\R$.  

If $S$ is a finite set, $\R[S]$ denotes the ring of polynomials in $S$ and $\R(S)$ denotes the extension field of $\R$ consisting of quotients of polynomials in $S$, that is, the  fraction field of $\R[S]$ (see e.g. \cite{Lang}). There is an evaluation map from $\R[S]$ to $\R$ obtained by assigning a real value to every element of $S$.

Let $\R_{+}$ denote the set of positive real numbers (excluding zero) and let  
$\overline{\R}_{+}$ denote the set of non-negative real numbers (including zero).
An increasing (resp. decreasing) continuous function $f(x)$ defined on an interval $I\subseteq \R$ is  
a continuous function such that $f(x)>f(y)$ (resp. $f(x)<f(y)$) for $x,y\in I$ satisfying $x>y$, i.e.~we consider strictly increasing/decreasing functions.

We will make several uses of the following result:
\begin{lemma}\label{rational} 
Let $f(x)=p(x)/q(x)$ be a rational function of the variable $x$, that is, a quotient of real polynomials $p(x)$ and $q(x)$ in $x$. Let $I\subseteq \R$ be an open interval. If $q(x)$ has no zeros in $I$, then $f(x)$ is a continuous function in $I$. 
In addition:
\begin{enumerate}[(i)]
\item If $f'(x)>0$ for all $x\in I$, then $f(x)$ is an increasing function in $I$. 
 \item If $f'(x)<0$ for all $x\in I$, then $f(x)$ is a decreasing function in $I$.
\end{enumerate}
If either $(i)$ or $(ii)$ are satisfied, then by the Inverse Function Theorem, $f$  admits a continuous inverse function $g\colon\im(f) \rightarrow  I$. The function $g$ is increasing (resp. decreasing) if  $f$ is.
\end{lemma}
If $q(x)$ is a real polynomial of degree $m$, then $q(x)$ has at most $m$ real zeros (it has exactly $m$ complex zeros when counted with multiplicity). If $x_1<\dots<x_r$ are the different real zeros of $q(x)$, ordered increasingly, then $f(x)$ is a continuous function in each of the $r+1$ intervals $(-\infty,x_1)$, $(x_r,+\infty)$ and $(x_i,x_{i+1})$ for $i=1,\dots,r-1$.

\subsection{One-site PTM signaling cascades}
\label{equations}

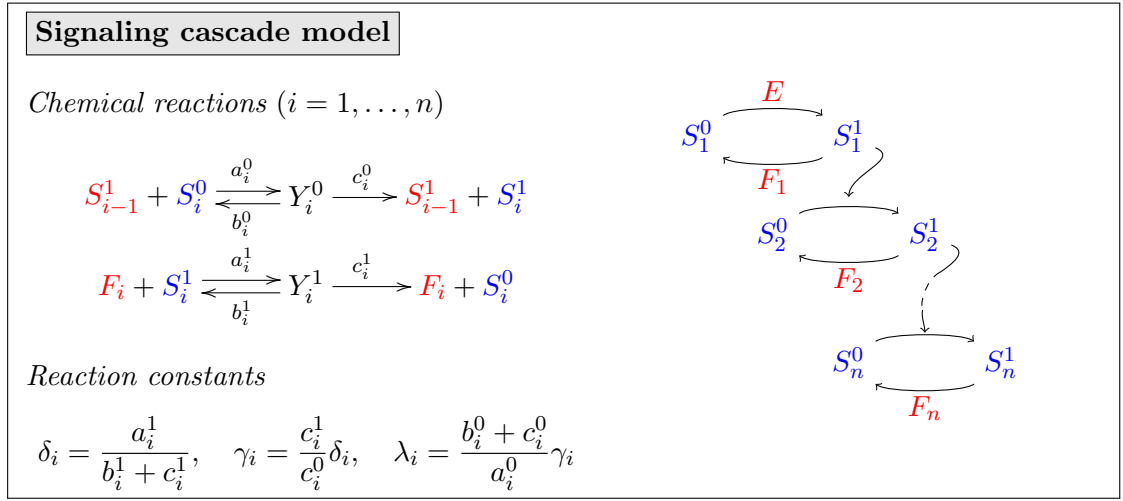
\begin{figure}[t]
\fcolorbox{black}{white}{
\begin{minipage}[h]{0.5\textwidth}
\fcolorbox{black}{white!80!gray}{\textbf{Signaling cascade model}}

\bigskip

{\it Chemical reactions} ($i=1,\dots,n$)
$$\xymatrix@R=15pt{
\textcolor{red}{S^{1}_{i-1}} + \textcolor{blue}{S^{0}_{i}} \ar@<0.5ex>[r]^(0.6){a_{i}^0} & Y^{0}_i \ar[r]^(0.35){c^0_i} \ar@<0.5ex>[l]^(0.4){b^0_i}& \textcolor{red}{S^{1}_{i-1}} +\textcolor{blue}{S^{1}_i} \\
\textcolor{red}{F_{i}}+\textcolor{blue}{S^{1}_i} \ar@<0.5ex>[r]^(0.6){a^{1}_i} & Y^{1}_i \ar[r]^(0.35){c^1_i} \ar@<0.5ex>[l]^(0.4){b^1_i}& \textcolor{red}{F_{i}}+ \textcolor{blue}{S_{i}^0} 
}$$

\textit{Reaction constants}
$$ \delta_{i}=\frac{a^{1}_i}{b^{1}_i + c^{1}_i}, \quad\gamma_{i}= \frac{c_i^{1}}{c_i^{0}} \delta_i,\quad \lambda_i=\frac{b^{0}_i + c^{0}_i}{a_{i}^0}\gamma_i $$

\end{minipage}
\begin{minipage}[h]{0.45\textwidth}
\centering
\begin{tikzpicture}
\node[blue] (S11) at (0,2)  {$S_{1}^{0}$};
\node[blue] (S21) at (2,2)  {$S_{1}^{1}$};
\draw[->] (S11) .. controls (0.5,2.4) and (1.5,2.4) .. (S21);
\draw[->] (S21) .. controls (1.5,1.6) and (0.5,1.6) .. (S11);
\draw[red] (1,2.6) node {$E$};\draw[red] (1,1.4) node {$F_{1}$};
\node[blue] (S12) at (1,0.7)  {$S_{2}^{0}$};
\node[blue] (S22) at (3,0.7)  {$S_{2}^{1}$};
\draw[->] (S12) .. controls (1.5,1.1) and (2.5,1.1) .. (S22);
\draw[->] (S22) .. controls (2.5,0.3) and (1.5,0.3) .. (S12);
\draw[->] (S21) .. controls (2.7,1.7) and (2,1.5) .. (2,1.2) ;
\draw[red] (2,0.1) node {$F_{2}$};
\node[blue] (S1n) at (2,-1)  {$S_{n}^{0}$};
\node[blue] (S2n) at (4,-1)  {$S_{n}^{1}$};
\draw[->] (S1n) .. controls (2.5,-0.6) and (3.5,-0.6) .. (S2n);
\draw[->] (S2n) .. controls (3.5,-1.4) and (2.5,-1.4) .. (S1n);
\draw (S22) .. controls (3.5,0.5) and (3.5,0.4) .. (3.2,0.2);
\draw[dashed] (3.2,0.2) .. controls (3,0.1) and (2.9,-0.3) .. (3,-0.3);
\draw[->] (2.97,-0.3) .. controls (2.95,-0.4) and (3,-0.5) .. (3,-0.6);
\draw[red] (3,-1.6) node {$F_{n}$};
\end{tikzpicture}
\end{minipage}

}
\caption{One-site PTM cascade of length $n$. The enzyme mechanism follows the classical Michaelis-Menten model. The reaction constants $\delta_i,\gamma_i,\lambda_i$ are introduced: $\delta_{i}$ is  the inverse of  the Michaelis-Menten constant for $F_i$, $\lambda_i$ is the relative catalytic efficiency in layer $i$, and $\gamma_i$ is the catalytic efficiency of $F_i$ divided by the dissociation constant of $S_{i-1}^1$, $c_i^0$.} \label{system}

\end{figure}

We consider signaling cascades with $n$ layers and one posttranslational modification (PTM) cycle at each layer, as illustrated in Figure \ref{system}. The chemical species involved in each cycle are the unmodified substrate $S_i^0$, the modified substrate $S_i^1$, the demodification enzyme $F_i$, the modification enzyme $E_i=S_{i-1}^{1}$ for $i=2,\dots,n$, and the intermediate enzyme-substrate complexes $Y_i^0$ and $Y_i^1$. 
That is, in each layer the modification enzyme is the activated substrate of the previous layer. The enzyme of the first layer, $E_1$, is not a substrate in any other layer. For convenience we put $S_{0}^1=E_{1}=E$. 
In the following,  signaling cascades of this kind are called  \emph{one-site PTM cascades}, or just cascades for short.

Although we consider PTM cycles of any type, we use the nomenclature of modification by phosphorylation and call the catalyzing enzymes for  \emph{kinase} (modification) and \emph{phosphatase} (demodification). 

The system is specified by the set of chemical reactions shown in Figure \ref{system}.
The enzyme mechanism  is assumed to follow the classical model of Michaelis and Menten, in which an enzyme-substrate complex is formed reversibly, while its dissociation into the product is considered irreversible. Further, the phosphate donor, which typically would be ATP, is assumed in constant concentration and not modeled explicitly. This reaction set-up is frequently used for studying signaling cascades; see  e.g.~\cite{Gold-Kosh-81,G-PNAS,bluthgen-switch,salazar_review,TG-rational,Ventura-Hidden}. 

Assuming mass-action kinetics, the differential equations describing the dynamics of the system over time $t$ are given by:

{\small
\begin{align}
\frac{dS^{1}_{i}}{dt} &= (b^{0}_{i+1}+c^{0}_{i+1}) Y^{0}_{i+1} +  c^{0}_{i} Y^{0}_{i} + b^{1}_{i} Y^{1}_{i} - (a^{0}_{i+1}S^{0}_{i+1} + a^{1}_{i}F_{i})S^{1}_{i}, \label{S1}\\
\frac{dS^{0}_{i}}{dt} &= b^{0}_{i} Y^{0}_{i} + c^{1}_{i} Y^{1}_{i} - a^{0}_{i}S^{0}_{i}S^{1}_{i-1},\label{S0}\\
\frac{dF_{i}}{dt} &= (b^{1}_{i}+ c^{1}_{i}) Y^{1}_{i} - a^{1}_{i}F_{i}S^{1}_{i},  \label{F} \\
\frac{dY^{0}_{i}}{dt} &=  -(b^{0}_{i}+c^{0}_{i}) Y^{0}_{i}  + a^{0}_{i}S^{0}_{i}S^{1}_{i-1},\label{Y0}\\ 
\frac{dY_{i}^1}{dt} &= -(b^{1}_{i}+ c^{1}_{i}) Y^{1}_{i} + a^{1}_{i}F_{i}S^{1}_{i},  \label{Y1}  
\end{align}}
\noindent
for $i=1\dots,n$. Here $a_i^*,b_i^*,c_i^*$ are positive rate constants. For convenience we put $c_{j}^*=b_{j}^*=a_{j}^*=0$ for $j=0, n+1$, such that equation \eqref{S1} also holds for $S^1_0=E$ (i.e. for $i=0$). The steady states are found by setting the  right hand side of the above equations  to zero, which results in a system of $5n+1$ polynomial equations  in $5n+1$ variables with coefficients in $\R$ (for given rate constants). However, \eqref{F} and \eqref{Y1} result in the same equation and similarly other equations are redundant because of the following \emph{conservation laws}:
\begin{equation}\label{totalamounts}
\overline{F}_i = F_{i} + Y^{1}_i, \quad \oE=E+Y_1^0,
\quad  \overline{S}_i  =   S^{0}_i + S^{1}_i + Y^{0}_i + Y^1_i + Y^{0}_{i+1}, 
\end{equation}
for $i=1,\dots,n$, and $Y_{n+1}^0=0$.  These can easily be verified by differentiation. The quantities $\oE$, $\oF_i$ and $\oS_i$ are called the total amounts of enzymes and substrates, or just the total amounts. 
Note that $\oE$ can be seen as the total amount of a $0$-th layer with only modified substrate.

From the form of the conservation laws we see that
  the steady state equations for $S_i^0$ and $Y_i^1$  (\eqref{S0} and \eqref{Y1}), and that corresponding to $E$ (\eqref{S1}, for $i=0$), are redundant. In addition, the steady state equation for $Y_{i+1}^0$  \eqref{Y0} implies that \eqref{S1}  can be rewritten as $c^{0}_{i} Y^{0}_{i} + b^{1}_{i} Y^{1}_{i} -  a^{1}_{i}F_{i}S^{1}_{i}=0,$ which together with \eqref{F} and \eqref{Y0} reduces to $c^{0}_{i} Y^{0}_{i} - c^{1}_{i} Y^{1}_{i} =0$. Therefore, the system of equations \eqref{S1}, \eqref{F} and \eqref{Y0} is equivalent to
\begin{equation*}\label{reduced1}
(a)\ c^{0}_{i} Y^{0}_{i} - c^{1}_{i}  Y_i^1  = 0,\quad
(b)\ Y^{1}_{i} - \frac{a^{1}_i}{b^{1}_i + c^{1}_i} F_{i}S^{1}_{i}  = 0,  \quad
(c)\  Y^{0}_{i}  -   \frac{a_{i}^0}{b^{0}_i + c^{0}_i} S^{1}_{i-1}S^{0}_{i} = 0,
 \end{equation*}
 for $i=1,\dots,n$. Further, this system is equivalent to the system  $\{(a)+c_i^1(b)$, $(b)$, $(a)+c_i^1(b)+c_i^0(c)\}$:
 \begin{equation}\label{reduced}
  Y_i^0-\gamma_i F_i S_i^1  =  0,\quad   Y^{1}_{i} - \delta_{i}F_{i}S_{i}^1= 0,\quad  \lambda_i F_i S_i^1  -S^{0}_{i} S_{i-1}^1  =   0 
\end{equation}
 for $i=1,\dots,n$, with the constants defined in Figure \ref{system}. The constant $\delta_{i}$ is  the inverse of  the Michaelis-Menten constant for $F_i$, $\lambda_i$ is the relative catalytic efficiency in layer $i$, and $\gamma_i$ is the catalytic efficiency of $F_i$ divided by the dissociation constant of $S_{i-1}^1$, $c_i^0$.

The system  consisting of \eqref{totalamounts} and \eqref{reduced} is called $\mathbf{SS}$ (for steady state) and the solutions to it are the  steady state concentrations for given total amounts. Let $\Con=\{\gamma_i,\delta_i,\lambda_i|\ i=1,\dots,n\}\cup \{\oS_i,\oF_i,\oE| \ i=1\dots,n\}$ be the set of \emph{parameters} of the system. Then $\mathbf{SS}$ is a system of polynomial equations with coefficients in $\R(\Con)$ and with variables being the chemical species of the cascade. It consists of  $2n$ quadratic equations (the last two equations in \eqref{reduced} for $i=1,\ldots,n$)  and $3n+1$ linear equations (the conservation laws \eqref{totalamounts} and the first equation in \eqref{reduced} for $i=1,\ldots,n$). 
We are going to show that for fixed total amounts, the steady states can be found from the zeros of a polynomial in one variable.

We note the following useful lemma:

\begin{lemma}\label{lemma_pos}
At any  steady state and for  $i=1,\ldots,n$,
\begin{itemize}
\item[(i)] If  $E=0$, then $\oE=0$.
\item[(ii)] If $F_i=0$, or $1+\delta_iS_i^1=0$, then $\oF_i=0$. 
\item[(iii)] If $S_i^1=0$, then either  $\oE=0$, or $\oS_j=0$ for some $j\leq i$.
\item[(iv)]  If $S_i^0=0$, then either  $\oF_i=0$, or $\oS_i=0$.
\end{itemize}
\end{lemma}
 If we allow $i=0$, then statement $(i)$ is included in $(iii)$.  From here onwards we assume the total amounts are positive.

By Lemma 2, $S_i^1=0$ and $1+\delta_iS_i^1=0$ are not solutions to the system $\mathbf{SS}$ for any  positive total amounts. Therefore, 
we can express the steady state concentrations of $F_i,Y_i^1,Y_i^0$, and $S_i^0$ in terms of those of the modified substrates:
\begin{equation}\label{Y0rel}
F_i = \frac{\oF_i}{1+\delta_iS_i^1},\quad Y_i^1=\frac{\delta_i \oF_i  S_i^1}{1+\delta_iS_i^1},\quad Y_i^0=\frac{\gamma_i\oF_i S_i^1 }{1+\delta_iS_i^1}, \quad  S_i^0=\frac{\lambda_i  \oF_i S_i^1}{(1+\delta_iS_i^1) S_{i-1}^{1}}.
\end{equation} 
The expression for $F_i$ is found from the second equation in \eqref{reduced} and the conservation law of $\oF_i$. The other equalities are obtained by substitution of this expression of $F_i$ in \eqref{reduced}.
We should remark here that the steady state values of $Y_i^0,Y_i^1$, and $F_i$ depend only on the rate constants in the $i$-th layer, and on $\oF_i$ and $S_i^1$. Additionally, they are  expressed as  rational functions of $S_i^1$. The steady state value of $S_i^0$, however, depends on the steady state values of the modified substrates in the $i$-th and $(i-1)$-th layers.

\subsection{Steady states as the zeros of a polynomial}\label{polysec}

Let $x_i=S_i^1$ for $i=0,1,\dots,n$, and $x_{-1}=1$. By substitution of the expressions of $Y_i^0,Y_i^1$ derived from \eqref{reduced} into $\oF_i,\oS_i$, the system $\mathbf{SS}$ is equivalent to the following system of equations:
\begin{eqnarray}
 Y_i^0-\gamma_i F_i x_i  =  0  \ \qquad \qquad  Y^{1}_{i} - \delta_{i}F_{i}x_{i}  &  =  & 0 \qquad i=1,\dots,n \label{simpl1} \\  \lambda_i F_i x_i  -S^{0}_{i} x_{i-1}  =   0  
   \qquad  (1+\delta_i x_i)F_i-\oF_i  &= & 0 \qquad i=1,\dots,n  \label{simpl2}\\
  S^{0}_i + x_i + (\gamma_i+\delta_i) F_i x_i + \gamma_{i+1} F_{i+1} x_{i+1} - \overline{S}_i  & = & 0 \qquad i=0,\dots,n \label{simpl3} \end{eqnarray}  
with the conventions  $\oS_0=\oE,\oF_{n+1}=0$, and $\delta_0=\gamma_0=\lambda_0=0$.
Since the system admits no solution of the form $x_{i}=0$ or $1+\delta_i x_i=0$  (Lemma \ref{lemma_pos}), we can eliminate the variables $F_i$ and $S_i^0$ in equation \eqref{simpl3} using \eqref{simpl2} to obtain an 
equivalent system  given by \eqref{simpl1}, \eqref{simpl2}, and the equation
 \begin{multline*}
 \lambda_i\oF_i (1+\delta_{i+1} x_{i+1}) x_i +  (1+\delta_i x_i)(1+\delta_{i+1} x_{i+1})x_{i-1}(x_i-\oS_i)  + \\  (\gamma_i+\delta_{i})\oF_{i}(1+\delta_{i+1} x_{i+1})x_{i-1}x_{i}  +  \gamma_{i+1} \oF_{i+1}(1+\delta_i x_i) x_{i-1}x_{i+1}=0,
\end{multline*}
 for $i=0,\dots,n$.
The left hand side of the equation can be written as a polynomial in $x_{i-1},x_i,x_{i+1}$ with coefficients in $\R(\Con)$,
 \begin{multline*}
q^S_i(x_{i-1},x_i,x_{i+1})= \\
(e^{1}_{i}x_{i+1} + e^{2}_{i})x_{i} +x_{i-1} (x_{i+1}(e^{3}_{i} x_{i}^2 + e_i^4 x_i + e^{5}_{i}) + e_i^6x_i^2 +e^{7}_{i}x_{i} + e^{8}_{i}),
 \end{multline*}
for $i=0,\dots,n$,  with coefficients: {\small $ e^{1}_{i}= \lambda_i\delta_{i+1}\oF_i,\ e^{2}_{i} =\lambda_i\oF_i, \ e^{3}_{i} = \delta_i\delta_{i+1}, \ e^{4}_{i} =   \delta_{i+1}e_i^7+ \gamma_{i+1}\delta_i\oF_{i+1}, \ e^{5}_{i} =  -\delta_{i+1}\oS_i + \gamma_{i+1}\oF_{i+1},\ e^{6}_{i} =  \delta_i,\  e^{7}_{i} =1+(\delta_i+\gamma_i)\oF_i - \delta_i \oS_i,$} and {\small $e^{8}_{i}= -\oS_i.$} For $i=0$, these reduce to {\small $e^4_0=\delta_{i+1},\ e^5_0=-\delta_{i+1}\oE +\gamma_{i+1}\oF_{i+1},\ e^7_0=1$,} and {\small $e^8_0=-\oE$}, with the remaining coefficients being equal to 0.

Therefore, the initial system, $\mathbf{SS}$, is equivalent to a system  consisting of two sets of equations: a system, $\mathbf{SS}_l$, of linear equations in $F_i,S_i^0,Y_i^0$ and $Y_i^1$ with coefficients in $\R(\Con \cup \{x_0,\dots,x_n\})$, corresponding to equations \eqref{simpl1} and \eqref{simpl2}, and a non-linear system,  $\mathbf{SS}_x=\{q_i^S=0\}$, of $n+1$ equations in $n+1$ variables, $x_i$, with coefficients in $\R(\Con)$.  

This procedure exemplifies a general property of the steady states of signaling cascades. Signaling cascades are studied in full generality in \cite{FAKW-1}, where we  extend previous results of Thomson and Gunawardena \cite{TG-rational}. We show that the steady states of a signaling cascade are determined by solving a system of $M$ polynomial equations (with coefficients depending on the reaction rates and total amounts) in $M$ variables. In our case $M=n+1$, the variables are $S_i^1$,  $i=0,\dots,n$, and the system is $\mathbf{SS}_x$.

Reciprocally, every solution to the system $\mathbf{SS}_x$ satisfying $x_{i}\neq 0$ and $1+\delta_i x_i \neq 0$ for all $i$, provides a solution to $\mathbf{SS}$. The values of the other steady state concentrations are found from the linear system $\mathbf{SS}_l$. In other words, if we let 
$$H(x_0,\dots,x_n)=\prod_{i=0}^{n} x_i (1+\delta_i x_i),$$
then we have proved that the steady states of the system are in one-to-one correspondence with the solutions to the system $\mathbf{SS}_x$ satisfying $H\neq 0$, that is, the solutions $x_i$ lying outside the hypersurface  $V_H:=\{H=0\}$.

\begin{remark}
One can see that for rate constants and total amounts in general position, all  solutions to the system $\mathbf{SS}_x$ lie outside $V_H$. Indeed, $x_i=0$ is a solution to the system if and only if   $h_1^i$ or $h_2^i$ vanish, where $h_1^i=(\gamma_{i+2}+\delta_{i+2})\oS_i - \lambda_{i+2}(\delta_{i+1}\oS_i - \gamma_{i+1}\oF_{i+1})$ and $h_2^i=\gamma_{i+3}+\delta_{i+3}-\delta_{i+2}\lambda_{i+3}$. The same conditions are obtained for solutions of the form $1+\delta_j x_j=0$. 
In this way, only if the parameter set fulfils $\prod_{i} h_1^i h_2^i=0$ has the smaller system $\mathbf{SS}_x$  non-valid solutions.
\end{remark}

The equations $q_i^S=0$ may be used iteratively to eliminate variables. Indeed, 
note that $q_i^S$ can be expressed in the form
$$ q_i^S(x_{i-1},x_i,x_{i+1})=a_{i}(x_i,x_{i+1}) + x_{i-1}  b_{i}(x_i,x_{i+1}),$$
 which is a linear polynomial in $x_{i-1}$ with coefficients in $\R(\Con \cup \{x_i,x_{i+1}\})$.  By Lemma \ref{lemma_pos}, the solutions to $\mathbf{SS}$ satisfy $1+\delta_ix_i,x_{i}\not=0$. Hence $a_i(x_{i},x_{i+1})=\lambda_i\oF_ix_i(1+\delta_{i+1}x_{i+1})\neq 0$, and  thus also  $b_{i}(x_i,x_{i+1}) \neq 0$. 
 
For $i=n$, we have $q_n^S(x_{n-1},x_n)=a_{n}(x_n) + x_{n-1}  b_{n}(x_n)$. This equation is used to eliminate the variable $x_{n-1}$ from $q_{n-1}^S=0$ to obtain a new equivalent equation  
$ \widetilde{q}_{n-1}^S(x_{n-2},x_n)=\widetilde{a}_{n-1}(x_n) + x_{n-2} \widetilde{b}_{n-1}(x_n)=0$,
satisfying $ \widetilde{a}_{n-1}, \widetilde{b}_{n-1}\neq 0$ for any solution to $\mathbf{SS}$.
We proceed iteratively in the same way for $i=n-2,\dots,1$ and obtain new equivalent equations,
\begin{equation}\label{rationalq}
\widetilde{q}_i^S(x_{i-1},x_n)= \widetilde{a}_{i}(x_n) + x_{i-1} \widetilde{b}_{i}(x_n)=0, 
\end{equation} 
which are linear polynomials in $x_{i-1}$ with coefficients in $\R(\Con \cup \{x_n\})$. Additionally, the coefficients are non-zero for any solution to $\mathbf{SS}$.

Finally, for $i=0$, we find the equation $q_0^S= (1+ \delta_1x_1)(x_0-\oE) + \gamma_1 \oF_1 x_1  =0.$  Elimination of $x_0$ and $x_1$  using $\widetilde{q}_1$ and $\widetilde{q}_2$, respectively, gives a polynomial in $x_n$:
\begin{equation}\label{polyq}
\widetilde{q}_0^S(x_n)=\left(\widetilde{b}_2(x_n)-\delta_1 \widetilde{a}_2(x_n)\right)\left(\widetilde{a}_1(x_n)+\oE \widetilde{b}_1(x_n)\right)+
\gamma_1\oF_1 \widetilde{b}_1(x_n) \widetilde{a}_2(x_n).
\end{equation}

To sum up, we have shown that any solution to the system $\mathbf{SS}$
provides a solution to the system $\widetilde{\mathbf{SS}}_x=\{\widetilde{q}_i^S=0\}$  together with $\mathbf{SS}_l $, and in particular that the steady state values of $x_n$ are  roots of the polynomial $\widetilde{q}_0^S(x_n)$.

The reverse might not be true depending on the  parameter values.  Specifically, a root of the polynomial $\widetilde{q}_0^S$ is a solution to $\mathbf{SS}$ if and only if $\widetilde{a_i},\widetilde{b}_i, \delta_{i-1}\widetilde{a}_i-\widetilde{b_i}\neq 0$.
 When this is the case,  the values of $x_0,\dots,x_{n-1}$ can be found from the root $x_n$ and $q_{i+1}^S$ as $x_{i}=-\widetilde{a}_{i+1}(x_n)/\widetilde{b}_{i+1}(x_n)$. The rest of the steady state concentrations are obtained by solving the linear system $\mathbf{SS}_l$ using \eqref{Y0rel}.

Note that equation \eqref{rationalq} gives $x_{i-1}=S_{i-1}^1$ as a rational function of $x_n=S_n^1$.
These functions will be studied in  detail in $\S$\ref{rationalsubstrates}.

\begin{theorem}[Polynomial]\label{poly} Consider a one-site PTM cascade with $n$ layers.
If positive total amounts  are given, then  any steady state value of $S_n^1$  is a root of the polynomial $\widetilde{q}_0^S(S_n^1)$. Additionally,  all  other  species concentrations  are expressed as rational functions of $S_n^1$. \end{theorem}

\begin{remark}
The elimination of variables from the system $\mathbf{SS}_x$  could have been started from the equation for $i=1$. This would have led to a polynomial in  $S_1^1$ describing the steady states of the system.
\end{remark}

\begin{remark}
The relation $F_i = \oF_i/(1+\delta_iS_i^1)$ gives $S_i^1=(\oF_i-F_i)/(\delta_i F_i)$. Using this relation, the steady state values of  $S_i^0,S_i^1,Y_i^0$, and $Y_i^1$ can be expressed as rational functions of $F_i$ instead. It follows that the steady states can be described by means of a polynomial in $F_1$ or $F_n$.
\end{remark}

We have proved that the steady states   are the roots of a polynomial. Additionally, we have shown that  the different variables are related through rational functions. Rational functions have quite well-described behaviors and this will enable us to derive theoretical results about the system. 

We are only interested in biologically relevant systems for which, in principle, it  is possible to form all substrates and intermediate complexes. This suggests the following definition: A \emph{Biologically Meaningful Steady State} (BMSS) is a steady state for which all total amounts are positive and all species concentrations are non-negative. 
Lemma \ref{lemma_pos} ensures that for a BMSS all species concentrations are  positive, i.e. non-zero. We call the set of BMSSs for  $D\subset \R_{+}^{5n+1}$. 

 It has been reported that multistationarity (more than one positive solution to the steady state equations  for fixed total amounts) occurs in the MAPK cascade \cite{Markevich-mapk,Qiao-bistability}. This cascade does not fit into our setting, because it has multiple site modification cycles in some layers. We will show in the following subsections that multistationarity cannot arise in our setting and that one-site PTM cascades have exactly one BMSS. 
 According to Theorem \ref{poly}, the value of $S_n^1$ determines the other concentrations at steady state. Therefore, multistationarity can only arise if the polynomial in $S_n^1$ has more than one zero for which all other  concentrations are in $D$. 
We will see that there is exactly one valid root, which is the first positive root of the polynomial. 

\subsection{Splitting the cascade into two parts}\label{breaking} 

We noted in the previous section that, provided positive total amounts are given, the steady state concentrations of a cascade of length $n$ are the solutions to the system $\mathbf{SS}$  given by \eqref{totalamounts} and  \eqref{reduced}.

Let us focus on the  cascade obtained from the first $i$ layers only. Its connection to the last $n-i$ layers is through the intermediate complex $Y_{i+1}^0$, accounting for the conversion of $S_{i+1}^0$ to $S_{i+1}^1$ via the kinase $S_i^1$. If the steady state value of $Y_{i+1}^0$ is known, then the steady state values of the chemical species in the first $i$ layers satisfy the steady state equations of a cascade of length $i$ with total amounts $\oE$, $\oF_1,\dots,\oF_i$, $\oS_1,\dots,\oS_{i-1}$, and $\oS_i-Y_{i+1}^0=S_i^0+S_i^1+Y_i^0+Y_i^1$.

Similarly for the cascade consisting of the last $n-i$ layers. If $S_{i}^1$ is known, then the steady state concentrations of the chemical species in layers $i+1,\dots,n$, satisfy the steady state equations of a cascade of length $n-i$ with total amounts $\oF_{i+1},\dots,\oF_n$ and $\oS_{i+1},\dots,\oS_n$.  

Therefore, the study of a cascade of length $n$ can be reduced to the study of smaller cascades. This observation is key in what follows and will be used to derive most of the conclusions about the cascades using induction arguments.

\subsection{The system with only one layer,  $n=1$} \label{onesteadyn1} 

We first focus on  understanding cascades of length $n=1$, that is, a system consisting of only one cycle. According to \eqref{Y0rel}, a steady state value of $S^1$ determines all other concentrations at steady state.

The system consists of  the chemical species $S^0, S^1,Y^0,Y^1,E$ and $F$.  Using \eqref{Y0rel}, the total amount $\oS$ in \eqref{totalamounts} (with the last term now being zero) is
\begin{equation}\label{sn1} 
\oS= \frac{\lambda  \oF S^1}{(1+\delta S^1) E} + S^1 + \frac{\gamma \oF S^1}{1+\delta S^1} + \frac{\delta \oF S^1}{1+\delta S^1}.\end{equation}
Further, use \eqref{totalamounts} and \eqref{Y0rel} to express $E$ in terms of $\oE$ and $S^1$, and conclude that   $\oS$ is  a rational function of $S^1$ of the form
\begin{align}\label{bS}
\overline{S} &=\varphi(S^1)=
\frac{S^1 p(S^1)  }{ q_{1}(S^1)q_{2}(S^1)},
\end{align}
where $\Delta=\gamma\oF-\delta\oE$, $q_{1}(S^1)=1 + \delta S^1$, $q_{2}(S^1)=\oE - \Delta S^1$, and $p(S^1)=\lambda\overline{F}q_{1}(S^1) + q_{1}(S^1)q_{2}(S^1) + (\gamma+\delta) \overline{F} q_{2}(S^1) $ is a polynomial of degree $2$. The function $\varphi$ can be extended continuously to $S^1=0$ for which  $\varphi(S^1)=0$. 

Let $\Gamma\subseteq \overline{\R}_{+}$ be defined as  $\Gamma=\overline{\R}_{+}$  if $\Delta\leq 0$, and 
$\Gamma=[0,\oE/\Delta)$ if $\Delta>0$.

\begin{proposition}\label{enzymebehavior} 
Let a one-site PTM cascade with $n=1$ layer and  positive total amounts be given. Then, 
\begin{enumerate}[(i)]
\item The system has a  unique BMSS. 
\item There exists an  increasing  continuous function $\psi:\overline{\R}_{+} \rightarrow \Gamma$   such that $S^1=\psi(\oS)$ is the BMSS value of $S^1$ and with $\psi(0)=0$.
\item Further, $\psi$  is the inverse of $\varphi$ on $\Gamma\subseteq \overline{\R}_{+}$. 
 \end{enumerate}
  \end{proposition}

Statement $(i)$ is well known and has been proven for instance in \cite{AngeliSontag,TG-Nature,WangSontag}. 
We note that $\varphi'$ is always positive and thus $\varphi$ is increasing in each of the intervals defined by the zeros of its denominator. Hence, it transpires that multiple steady states could only occur if  $\Delta>0$.  The previous proposition ensures that in this case the BMSS is located between zero and $\oE/\Delta$ and values of $S^1$ larger than $\oE/\Delta$ result in  biologically non-valid concentrations. 

We now introduce  another rational function relating $E$ to $Y^0$.
If we isolate $E$ from   \eqref{sn1},  we obtain $E$ as a function of $S^1$ and this function  does not depend on $\oE$.  Substituting  $S^1=Y^0/(\gamma \oF - \delta Y^0)$ (obtained from \eqref{Y0rel}) into this expression, we get the following relation at steady state,
\begin{equation}\label{en1}
E=g(Y^0)=\frac{p_1(Y^{0})}{p_2(Y^{0})},
\end{equation}
where
$$
p_1(y)  =  \lambda y(\xi- y), \quad
p_2(y) = (\delta+\gamma)y^2 - \gamma (1/\delta +\oF+\xi + \oS)y + \gamma\xi\oS,$$
 and  $\xi=\gamma\oF/\delta$.

 \begin{proposition}\label{functiong} The polynomial $p_2$ has two positive roots  $\alpha_1$ and $\alpha_2$ with $\alpha_1<\xi<\alpha_2$ and  the function $g$ is increasing and continuous in the three intervals defined by these two roots. The BMSS  exists  for values of $Y^0$ in $(0,\alpha_1)$ only. Additionally,  $g(0)=0$, and $g(Y^0)$ tends to $+\infty$ as $Y^0$ tends to $\alpha_1$.
 \end{proposition}

It follows that the BMSS value of $E$ is given as a  continuous increasing function of the BMSS value of $Y^0$. A choice of $\oE=E+Y^0=g(Y^0)+Y^0$ imposes the BMSS value of $Y^0$.

\subsection{One-site PTM cascades admit exactly one BMSS}\label{onesteady}

To proceed to higher values of $n$, we use an induction argument. Given a cascade of length $n$ we consider separately  the last layer  $C_{n}$ of the cascade and the cascade $C$ consisting of the first $n-1$ layers. According to $\S$\ref{breaking},  the link between the two cascades is  the complex $Y^{0}_{n}$.  

If we focus on the last layer $C_n$, the kinase of this system is  $S^{1}_{n-1}$. Recall that the function $g_n=g$ defined  in \eqref{en1} does not involve $\oE$. Thus, for  any BMSS, the relation
\begin{equation*}
S_{n-1}^1 = g_n(Y^0_n)\end{equation*}
holds  with constants $\gamma_n,\delta_n,\lambda_n,\oF_n,$ and $\oS_n$. By Proposition \ref{functiong}, $g_n$ is a continuous increasing function in $[0,\alpha_1)$ with image  $\overline{\R}_{+}$.

We focus now on the cascade $C$. As noted previously, $C$ corresponds to a cascade of length $n-1$  with total amounts  $\overline{E},\overline{F}_1,\dots,\overline{F}_{n-1},\overline{S}_1,\dots,\overline{S}_{n-1}-Y^{0}_n$.  The key to proceed  is  that for a cascade of length $n-1$, $S^{1}_{n-1}$ is an increasing  continuous function $\psi_{n-1}$ of $\overline{S}_{n-1}$ with $\psi_{n-1}(0)=0$  (this is proved by induction in Theorem \ref{onesteadystate}). Therefore, since the total amount of substrate in the last layer of $C$ is $\oS_{n-1}-Y_n^0$, we have that 
$$S_{n-1}^1=f(Y_n^0) := \psi_{n-1}(\oS_{n-1}-Y_n^0),$$
with $f$ a decreasing function in $[0,\oS_{n-1})$.

\begin{figure}[t]
\framebox[\textwidth]{
\begin{minipage}[!tbp]{0.95\textwidth}
\begin{minipage}[!tbp]{0.5\textwidth}
\fcolorbox{black}{white!80!gray}{\textbf{Determination of the BMSS}}

\medskip 
\begin{tikzpicture}
\draw[fill=blue,very nearly transparent,rounded corners] (-0.3,2.9) rectangle (3.5,-0.5);
\draw[draw=blue,line width=1pt,rounded corners] (-0.3,2.9) rectangle (3.5,-0.5);
\node[anchor=west,text width=2.4cm,text badly centered] (t1) at (3.6,1.3) {$S_{n-1}^1=f(Y_n^0)$ as substrate};
\node[anchor=west,blue, text width=2.4cm,text badly centered] (t1) at (3.6,0.6) {(blue curve)};
\draw[fill=red,very nearly transparent,rounded corners] (1.7,-0.6) rectangle (4.3,-1.8);
\draw[draw=red,line width=1pt,rounded corners] (1.7,-0.6) rectangle (4.3,-1.8);
\node[anchor=north west,text width=2.5cm,text badly centered] (t1) at (-1,-0.6) {$S_{n-1}^1=g_n(Y_n^0)$ as kinase};
\node[anchor=north west,red, text width=2cm,text badly centered] (t1) at (-0.75,-1.6) {(red curve)};

\node (S11) at (0,2)  {$S_{1}^{0}$};
\node (S21) at (2,2)  {$S_{1}^{1}$};
\draw[->] (S11) .. controls (0.5,2.4) and (1.5,2.4) .. (S21);
\draw[->] (S21) .. controls (1.5,1.6) and (0.5,1.6) .. (S11);
\draw (1,2.6) node {$E$};
\draw (1,1.4) node {$F_{1}$};
\node (S12) at (1,0.3)  {$S_{n-1}^{0}$};
\node[white,circle,nearly transparent,fill=yellow,inner sep=0.5pt] (S22) at (3,0.3)  {$S_{n-1}^{1}$};
\node (S22) at (3,0.3)  {$S_{n-1}^{1}$};
\draw[->] (S12) .. controls (1.5,0.7) and (2.5,0.7) .. (S22);
\draw[->] (S22) .. controls (2.5,-0.1) and (1.5,-0.1) .. (S12);
\draw (2,-0.3) node {$F_{n-1}$};
\node (S1n) at (2,-1)  {$S_{n}^{0}$};
\node (S2n) at (4,-1)  {$S_{n}^{1}$};
\draw[->] (S1n) .. controls (2.5,-0.6) and (3.5,-0.6) .. (S2n);
\draw[->] (S2n) .. controls (3.5,-1.4) and (2.5,-1.4) .. (S1n);
\draw (3,-1.6) node {$F_{n}$};
\draw[-] (S21) .. controls (2.7,1.8) and (2,1.6) .. (1.95,1.32) ;
\draw[dashed,->] (1.95,1.3) .. controls (1.9,0.9) and (2,0.9) .. (2,0.7);
\draw[->] (S22) .. controls (3.8,-0.2) and (2.9,-0.2) .. (3,-0.6);

\end{tikzpicture}
\end{minipage}
\hspace{0.3cm}
\begin{minipage}[!tbp]{0.45\textwidth}
\begin{flushright}
\includegraphics[scale=0.5]{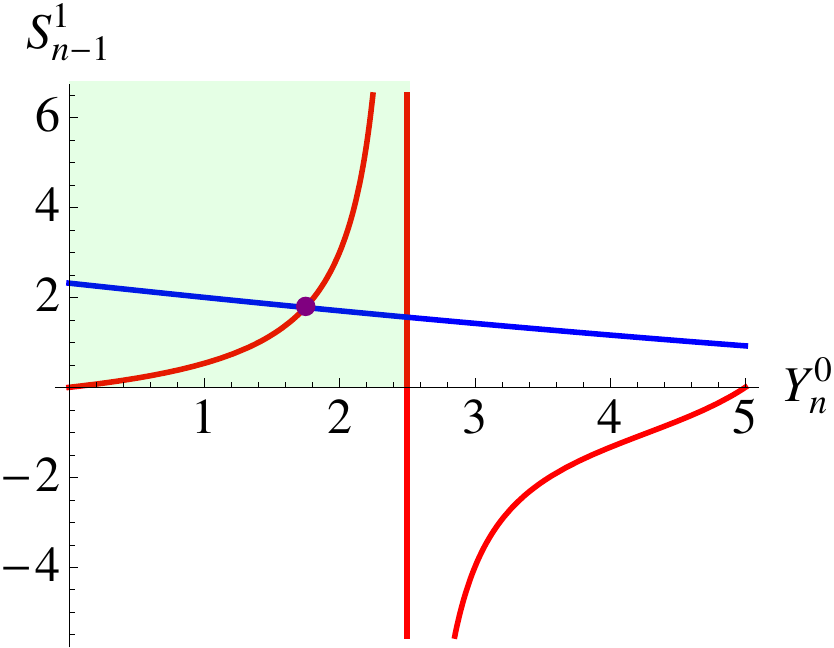}
\end{flushright}
\medskip

\footnotesize{BMSSs are located in the highlighted region (before the first singularity of $g_n$). The two curves intersect in the BMSS value of $(Y_n^0,S_{n-1}^1)$.}
\end{minipage}
\end{minipage}
}
\caption{Induction argument to show the existence of exactly one BMSS}\label{oneintersect}
\end{figure}
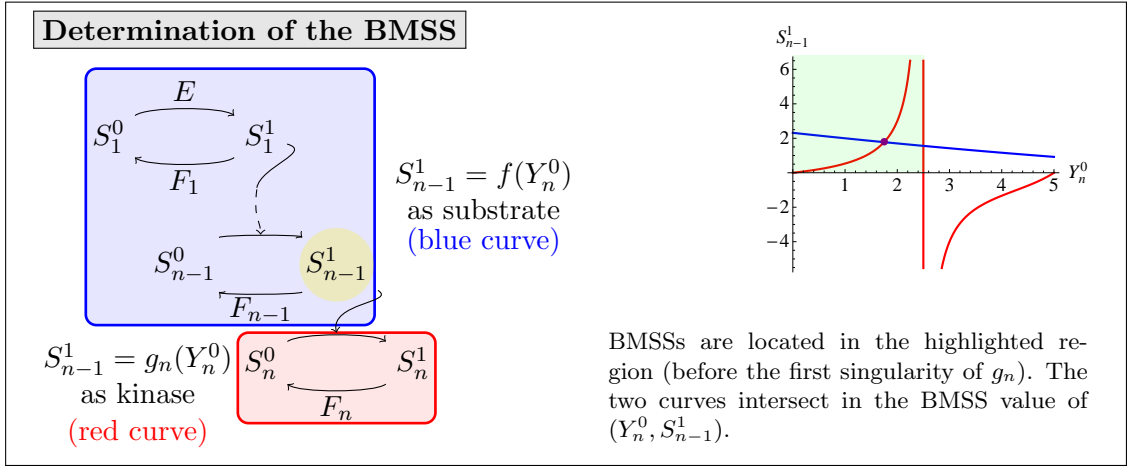

Thus, the split of the cascade provides two functions of $Y_n^0$ describing the BMSS value of $S_{n-1}^1$.
Both functions are continuous, $g$ is increasing and tends to infinity as $\alpha_1$ is approached, while $\psi_{n-1}$ is decreasing. It follows that they intersect in exactly one point, which is the BMSS value of $Y_n^0$. This is illustrated in Figure \ref{oneintersect}. The technical details of this argument are covered in the proof of Theorem \ref{onesteadystate}. 

Let 
$\Delta_n=\gamma_n \oF_n-\delta_n \oS_{n-1}=\delta_n(\xi_n-\oS_{n-1})$ with $\xi_n=\gamma_n\oF_n/\delta_n$ 
such that  $\Delta_1=\Delta$ and $\xi_1=\xi$. Further, let $\Gamma_n=\overline{\R}_{+}$ for $\Delta_n\leq 0$, and $\Gamma_n=[0,\oS_{n-1}/\Delta_n)$ for $\Delta_n>0$, such that $\Gamma_1=\Gamma$.

 \begin{theorem}[One steady state]\label{onesteadystate}
Let a one-site PTM cascade with  $n\geq 1$ layers and  positive total amounts be given. Then, the cascade has a  unique steady state  in $D$. Further, if $\oS_n$ is considered variable, then
 \begin{enumerate}[(i)]
\item $Y_n^{0}$ is  a  continuous increasing function of $\oS_n$ with image $[0,\mu_n)$, where $\mu_n=\min(\xi_n,\oS_{n-1})$.
\item $S^{1}_{n}$ is an increasing  continuous function $\psi_n$ of $\overline{S}_{n}$ with $\psi_n(0)=0$. It is the inverse of a rational function $\varphi_n$ defined on $\Gamma_n$.  For $n=1$, $\varphi_1=\varphi$ and $\psi_1=\psi$.
\end{enumerate}
\end{theorem}

\begin{remark}
Our approach is not only useful for  showing that there is exactly one BMSS. It also provides a constructive and iterative way of finding it. A priori we have no explicit analytical expression for $\psi_n$, but we do for its inverse $\varphi_n$.
\end{remark}

\begin{remark}[Effective computation] The rational function $\varphi_n$ and the set $\Gamma_n$ can  be determined from the total amounts and the rate constants. Programs like Mathematica\texttrademark{} allow for the effective computation of an inverse function $x=g(y)$ of $y=f(x)$ using the command \emph{Interpolate}. It proceeds from a list of pairs $(y,g(y))=(f(x),x)$. By generating many pairs $(f(x),x)$, the estimated inverse function can be as accurate as desired.
\end{remark}

\begin{remark}[Stability of the BMSS] \label{stability}
We have shown that a one-site PTM cascade admits exactly one BMSS for any set of rate constants and any specified positive total amounts. This steady state is biologically attainable if it is (asymptotically) \emph{stable}, that is, nearby trajectories are attracted to it.
For a system of ordinary differential equations, a steady state is asymptotically stable if all  eigenvalues of the Jacobian have negative real parts \cite[Thm. 1.1.1]{Wiggins}. For $n=1$, it is shown in \cite{AngeliSontag} that the steady state is a global attractor. 
For $2\leq n\leq 10$, we  randomly generated parameters ranging from $10^{-6}$ to $10^4$. In all cases we have found that the stability criterion is satisfied and hence that the BMSS is stable. However, we have been unable to provide a mathematical proof. \end{remark}

 \subsection{Rational functions relating the substrates}\label{rationalsubstrates}

In this subsection we are going to take a closer look at the rational functions relating the modified substrates derived from equation \eqref{rationalq}. We will provide an iterative expression for them that  elucidates different biologically relevant properties.

For convenience, define the rational function  relating $Y_i^0$ and $S_i^1$ in \eqref{Y0rel} by 
\begin{align}\label{gy}
Y_i^0 & =g_i^Y\!(S_i^1)=\frac{\gamma_i  \oF_i S_i^1}{1+\delta_iS_i^1}.
\end{align}
Consider a  cascade of length $n$. As we have seen, steady state concentrations in the last layer satisfy the relation $S_{n-1}^1=g_n(Y_n^0)$. 
 Combined with \eqref{gy}, this gives the relation
\begin{equation*}
S_{n-1}^1 = f_{n-1}(S_n^1)=\frac{ \lambda_n\oF_n S_n^1}{d_n(S_n^1,0)},
\end{equation*}
 with
 \begin{equation*}
d_i(x,y)= (\oS_i-y) - x - \oF_i( \delta_i +\gamma_i) x + \delta_i(\oS_i-y) x - \delta_i x^2,
\end{equation*}
$1\leq i\leq n$. (The reason for introducing $d_i$ as a function of two variables will be clear later.)
In $d_n(x,0)$, the sign of the leading coefficient divided by the sign of the independent term is negative, hence the polynomial has exactly one positive real root, say $\alpha_n$. For $x=0$, we have $d_n(x,0)>0$. Consequently, $S_{n-1}^1$ is only meaningful for $S_n^1\in [0,\alpha_n)$ (is negative otherwise). 
Therefore, $\oF_n$, $\oS_n$ and the rate constants of layer $n$ set an upper bound to $S_n^1$ for any steady state, independently of the parameters in the previous layers of the cascade. In fact, the previous layers further constrain $S_n^1$, e.g. $\oS_{n-1}$ establishes an upper bound to $S_{n-1}^1$ and hence the maximal value of $S_n^1$ is further decreased.  (Here and elsewhere, `maximal value' is the smallest upper bound, i.e. the supremum, but it is not attained.)

In general, consider the $i$-th layer of the cascade. 
For every value of $Y_{i+1}^0$, the steady state values of the species in the first $i$ layers are found by solving the steady state equations for the cascade consisting of layers $1$ to $i$  with the total amount of substrate in layer $i$ being $\oS_i-Y_{i+1}^0$. Therefore, applying the same argument as above, we have
\begin{equation}\label{rationalsi}
S_{i-1}^1=g_{i}(S_i^1,Y_{i+1}^0)=  \frac{ \lambda_i\oF_i S_i^1}{d_i(S_i^1,Y_{i+1}^0)},
\end{equation}
with $g_{n}(S_n^1,Y_{n+1}^0)=f_{n-1}(S_n^1)$, since $Y_{n+1}^0=0$. 
Arguing again as above, the polynomial $d_i(S_i^1,Y_{i+1}^0)$  has exactly one positive real root in $S_i^1$, say $\rho_i(Y_{i+1}^0)$,  for a fixed value of $Y_{i+1}^0\in [0,\oS_i)$. 

Consider $0<y_1<y_2$. Writing $y_2=y_1+(y_2-y_1)$, we find that
\begin{align*}
d_i(\rho(y_1),y_2) & =  -(y_2-y_1) -\delta_i(y_2-y_1)\rho(y_1) <0
\end{align*}
and therefore $\rho(y_2)<\rho(y_1)$. In particular,  if $\alpha_i=\rho_i(0)$, then $\rho_i(Y_{i+1}^0)\leq \alpha_i$ and  $S_i^1<\alpha_i$ at any steady state.

It follows that for every $i$, there is an upper bound, $\alpha_i$, for $S_i^1$ at steady state, depending only on the rate constants and the total amounts, $\oF_i$ and $\oS_i$, in the layer. How far the steady state value is from the upper bound depends on  the amount of sequestered substrate in $Y_{i+1}^0$.

\begin{proposition}[Rational functions]\label{rationalfunctions}
For  $i=0,\ldots,n-1$,  the BMSS value of $S_i^1$  satisfies $S_i^1=f_{i}(S_n^1)$, where $f_i$ is an increasing   rational function of $S_n^1$ defined on an interval $[0,\beta_i)$. It depends on $\oF_j, \oS_j$, $j\geq i+1$. Furthermore,  $\beta_i<\beta_{i+1}$ and $\beta_i<\beta_{n-1}=\alpha_n$ for $i<n-1$. 
In fact, the following relation holds
\begin{equation*}
S_{i}^1= f_{i}(S_n^1)= \frac{ \lambda_{i+1}\cdots \lambda_n\oF_{i+1}\cdots \oF_n S_n^1}{d_{i+1}(S_{i+1}^1,Y_{i+2}^0)\cdots d_n(S_n^1,0)},
\end{equation*}
where  $S_j^1=f_j(S_n^1)$, $j\geq i+1$ and $Y_j^0=f_j^Y(S_n^1)=g_j^Y(f_j(S_n^1))$, $j\geq i+2$, are given recursively via  \eqref{gy} and \eqref{rationalsi}.
In addition,
\begin{enumerate}[(i)]
\item $\beta_i$ is the first positive singularity of $f_i$ and a root of $d_{i+1}(x,Y_{i+2}^0)$. In particular, the image of $f_i$ over $[0,\beta_i)$ is $\overline{\R}_{+}$. 
\item If $\oS_{i+1}$ tends to $+\infty$, then $\beta_i$ tends to $\beta_{i+1}$ (with $\beta_n=+\infty$).
\item $S_i^1<\alpha_i$.
\end{enumerate}
\end{proposition}

\begin{figure}[t]
\framebox[\textwidth]{

\begin{minipage}[!tbp]{0.98\textwidth}
\hspace{0.4cm}
\begin{minipage}[!tbp]{0.9\textwidth}
\fcolorbox{black}{white!80!gray}{\textbf{Rational functions}}

\medskip
We illustrate the rational functions and bounds in Proposition \ref{rationalfunctions} for a cascade of length $n=3$.
Graphics (a-c) below depict the rational functions $f_i$ relating the steady state values of $S_2^1,S_1^1,E$ to $S_3^1$. The first singularity, $\beta_i$, is highlighted in orange and delimits the region for the BMSS (in green). The a priori maximal value $\alpha_i$ (the positive root of $d_i(x,0)$\ ) is also depicted. Note that $\alpha_i$ reduces the possible maximal value of $S_3^1$. The highlighted region in graphic (d) corresponds to the inverse of the stimulus-response curve, see $\S$\ref{stimresponse}.\end{minipage}

\vspace{0.3cm}

\begin{minipage}[!tbp]{0.47\textwidth}
\subfloat[$S_2^1=f_2(S_3^1)$]{
\includegraphics[scale=0.65]{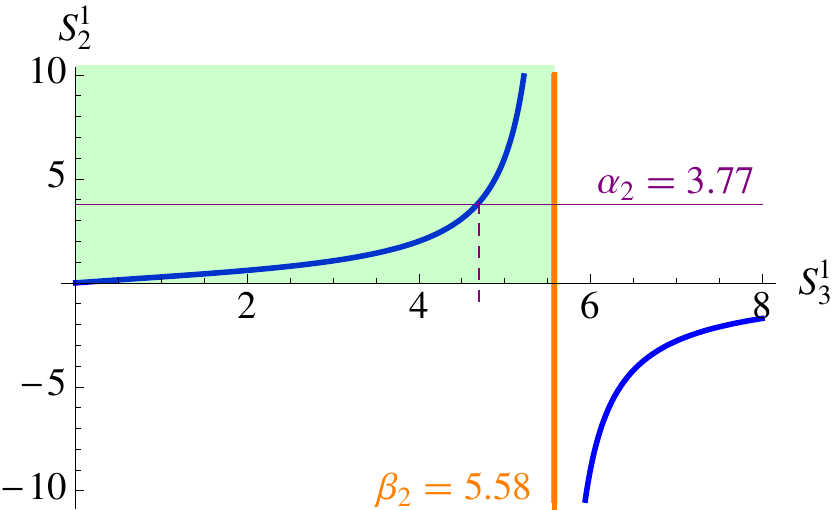}}
\end{minipage}
\hspace{0.4cm}
\begin{minipage}[!tbp]{0.45\textwidth}
\subfloat[$S_1^1=f_1(S_3^1)$]{
\includegraphics[scale=0.65]{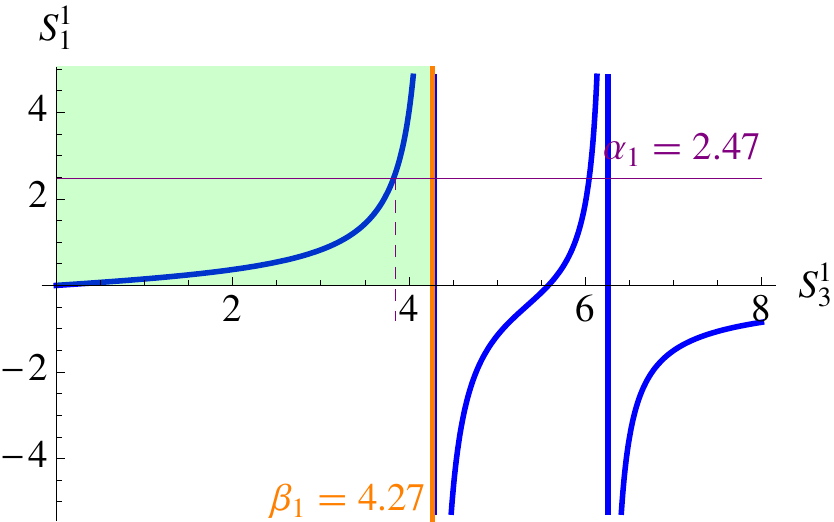}}
\end{minipage}

\begin{minipage}[!tbp]{0.47\textwidth}
\subfloat[$E=f_0(S_3^1)$]{
\includegraphics[scale=0.65]{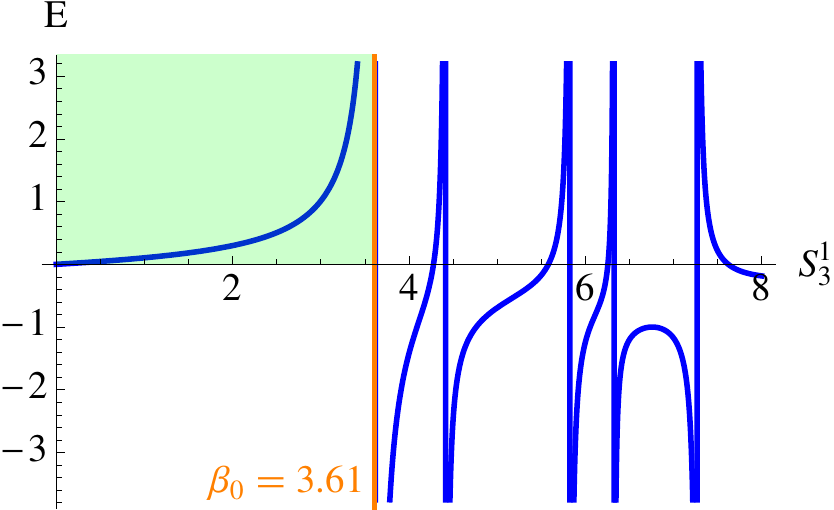}}
\end{minipage}
\hspace{0.4cm}
\begin{minipage}[!tbp]{0.45\textwidth}
\subfloat[$\oE=E+Y_1^0=r(S_3^1)$]{
\includegraphics[scale=0.65]{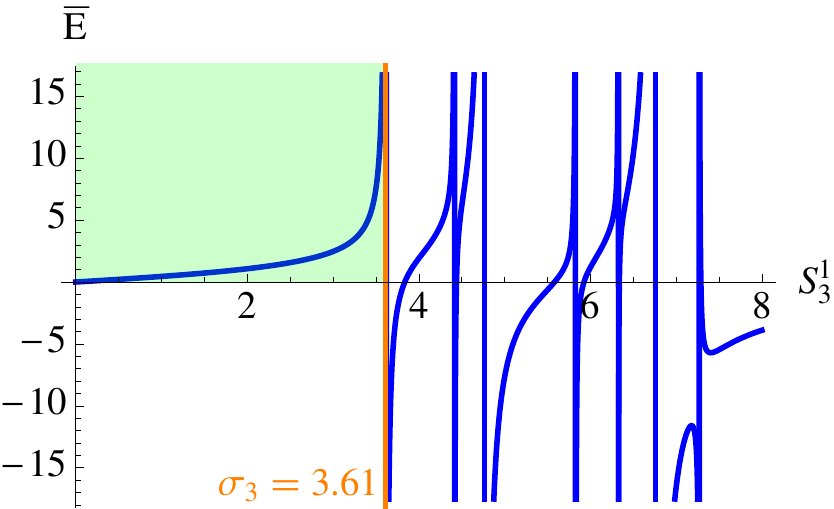}}
\end{minipage}

\vspace{0.4cm}
\begin{minipage}[h]{0.9\textwidth}
\tiny{Parameters: \hspace{0.36cm}
$a_*^*=b_*^*=c_*^*=1$, $ \oF_1=5,\oF_2=4,\oF_3=3$, $\oS_1=8,\oS_2=9,\oS_3=10$
\\
Polynomials: \hspace{0.3cm} $d_1(x,0) =8-2x-0.5x^2$, $d_2(x,0)=9-0.5x-0.5x^2$, $d_3(x,0)=10+x-0.5x^2$
}
\end{minipage}
\end{minipage}
}
\caption{Rational functions for the modified substrate and stimulus-response.}\label{ratfig}
\end{figure}

In Figure \ref{ratfig}, we show the graphics of the rational functions, and the values $\beta_i$ and $\alpha_i$  for a cascade of length $n=3$.

The rational function for $S_{i-1}^1$ in terms of $S_n^1$ is obtained by considering $S_{i-1}^1$ as the enzyme in the $i$-th layer. Therefore, we require layers $2,\dots,n$ to compute the rational functions of $S_1^1,\dots,S_{n-1}^1$ in terms of $S_n^1$, and layer $1$ to compute that of $E=S_0^1$. Hence they are independent of $\oE$. When we require the rational functions to satisfy the conservation law  for $\oE$,  the steady state value of $S_n^1$ is determined  and those of the other variables can subsequently be derived using  \eqref{Y0rel} and Proposition \ref{rationalfunctions}. 

From these results, one  derives the  following theorem.

\begin{theorem}[Downstream variability]\label{down}  Assume a one-site PTM cascade of length $n$ is given with
all total amounts but $\oS_i$, for some $i$,  fixed.   Then, increasing $\oS_i$ causes the steady state values of $S_j^1,Y_j^0,$ and $Y_j^1$ to increase for all $j=i,\dots,n$. 
\end{theorem}

By Theorem \ref{onesteadystate}(ii), if $\oS_n$ increases or decreases, then so does $S_n^1$. Thus, if we consider the first $j$ layers of the cascade, then an increase or decrease of the total amount $\oS_j-Y_{j+1}^0$ will induce the same effect on  $S_j^1$. On the other hand, by Theorem \ref{down}, if  all total amounts but $\oS_i$ are fixed, an increase in  $\oS_i$ causes the steady state value of  $Y_{i}^0$ to increase and therefore $\oS_{i-1}-Y_i^0$ to decrease. It follows that $S_{i-1}^1$ and $Y_{ i-1}^0$ decrease (Theorem \ref{down}). Proceeding with the same reasoning,  we see that the effect is transmitted upstream in an alternating fashion.
  
\begin{theorem}[Upstream variability]\label{alternation} Assume a one-site PTM cascade of length $n$ is given with
all  total amounts but $\oS_i$, for some $i$,  fixed. Then, increasing $\oS_i$ causes the steady state values  of $S_j^1$, $j=1,\ldots,i$, to  increase for $j=i-2k$, and to decrease for $j=i-2k-1$.
\end{theorem}

 \subsection{Stimulus-response curves as rational functions}\label{stimresponse}

One-site PTM cascades have exactly one BMSS for any initial total amounts. Hence,  for every value of $\oE$ and all other  total amounts fixed, there is a unique steady state value of $S_n^1$. The plot of $S_n^1$ against $\oE$ is usually called the \emph{stimulus-response} curve. 

By  \eqref{gy} and \eqref{rationalsi} we have
$$\oE = E + Y_1^0 = g_1(S^1_1,Y_2^0) + g_1^Y(S_1^1),\ \textrm{with }\begin{cases} g_1(S^1_1,Y_2^0)=\frac{ \lambda_1\oF_1 S^1_1 }{d_1(S^1_1,Y_2^0)} \\ g_1^Y(S_1^1)= \frac{\gamma_1\oF_1 S^1_1}{1+\delta_1 S^1_1}.\end{cases}$$

Note that $f_1^Y=g_1^Y\circ f_1$ is a  continuous increasing function of $S^1_n\in [0,\beta_1)$ (Proposition \ref{rationalfunctions}). 
As a function of $S_n^1$, $g_1$ is $f_0$ and hence by Proposition \ref{rationalfunctions}, it is a  continuous increasing function of $S_n^1\in [0,\beta_0)$. Since $\beta_0<\beta_1$,  we obtain the following theorem.

\begin{theorem}[Stimulus-response curves]\label{stimulus}
The stimulus-response curve $S_n^1=sr(\oE)$ is the inverse of a rational function
$$\oE=r(S_n^1)=\frac{r_1(S_n^1)}{r_2(S_n^1)}=f_0(S_n^1)+ f_1^Y(S_n^1),$$
for $S_n^1$ in $[0,\sigma_n)$,  where $\sigma_n=\beta_0$ is the first positive real zero of $r_2(S_n^1)$. The polynomials $r_1$ and $r_2$ depend only on the parameters of the system and the specified total amounts of substrates and phosphatase (not the kinase). In addition, $r$ is  continuous and increasing.
\end{theorem}

The plot in Figure \ref{ratfig}(d) shows that $r$ might have  many singularities. However, the biologically meaningful region is given as the interval from zero to the first singularity.

By construction, the polynomial $\widetilde{q}_0^S(S_n^1)$ in \eqref{polyq} is exactly $r_2(S_n^1)\oE - r_1(S_n^1)$ (up to multiplication by a real number). Therefore, we obtain the following corollary.

\begin{corollary}\label{firstroot} Let a one-site PTM cascade with $n$ layers and positive  total amounts be given. The  BMSS value of $S_n^1$ is  the first positive real root of the polynomial $\widetilde{q}_0^S$ in $S_n^1$, as given in Theorem \ref{poly}.
\end{corollary}

To finish the subsection we note the following. With fixed total amounts of substrates and phosphatase, all steady state concentrations are given as rational functions of $S_n^1$, independently of $\oE$. This implies that the stimulus-response curve showing the response of any species in the cascade admits a  \emph{rational algebraic parameterization} in the form
\begin{eqnarray*}
 r_C: [0,\sigma_n) & \rightarrow  & \overline{\R}^2_{+} \\ 
 s & \mapsto & (r(s),f_C(s)),
 \end{eqnarray*}
where $C$ is a chemical species and  $f_C(S_n^1)$ is the corresponding rational function of $S_n^1$ for that species. The maximal values of $S_1^1,\dots,S_{n-1}^1$ are obtained as $\sigma_i=f_{i}(\sigma_n)$.

\section{Biological implications}
\label{biol_sec}

In  section $\S$\ref{math_sec} we have focused on the analytical description of one-site PTM cascades of arbitrary length $n$. 
The existence of exactly one biologically meaningful steady state has been shown. Additionally, our approach has provided explicit rational functions  relating   concentrations of substrates and enzymes at steady state. 

In this section we exemplify how our method can be used to provide qualitative insight into different biological aspects of signaling cascades.
Generally, it is difficult to obtain experimental data and to estimate rate constants. Rate constants are typically only known  in  specific experimental contexts or to be within a certain range. It is therefore of importance to be able to derive conclusions that do not stringently rely on specific values of rate constants \cite{perspective,shinar}.
We show that this is possible using our approach and that different cascade behaviors, for instance in response to noise or varied stimuli, can be studied using our analytical description.  Further, it might be possible to design or guide experiments from the expected behavior of a system  in order to  uncover missing connections  in a reaction network or the presence of feedback mechanisms \cite{G-PNAS,G-distributivity,shinar}. 

In what follows, stimulus refers to $\oE$ while response refers to the steady state value  $S_n^1$ of the modified substrate in the last layer.

\subsection{Maximal response}

The maximal response of a cascade is the limiting steady state value of $S_n^1$, when the stimulus $\oE$ is increased to infinity, or in realistic terms, it is the value of the response $S_n^1$, when the stimulus is  large. The total amount of substrate in the last layer sets an obvious upper bound to the maximal response. However, the maximal response can be far from the total amount of substrate. The first bound is imposed by the rate constants and the level of  phosphatase in the last layer (by $\alpha_n$, cf. $\S$\ref{rationalsubstrates}). The bound is further restricted  by the amount of substrate in the previous layer (since the substrate acts as kinase), and, as we will see below, the length of the cascade.

We have shown that the maximal response attainable in a cascade of length $n$ is given by the first positive zero of a polynomial, which corresponds to the denominator of the function $r$ (Theorem~\ref{stimulus}). In  Proposition \ref{rationalfunctions} we saw  that every layer of the cascade accounts for a reduction of the maximal response. In this sense, adding a new layer on top of the cascade lowers the possible maximal response. Only if the total amount in the new layer is very large, will the maximal response  remain unchanged or essentially the same, cf.~Proposition \ref{rationalfunctions}(ii). 
 
 \begin{figure}[t]
\framebox[\textwidth]{

\begin{minipage}[!tbp]{0.44\textwidth}
\fcolorbox{black}{white!80!gray}{\textbf{Maximal response}}

\medskip 
Reduction of the maximal value of response $S_n^1$ for different cascade lengths $n$ (for fixed parameters in the last layer)
\begin{center}
\begin{tabular}{cc}
Length &  Max ($\sigma_n$)
\\ \hline
1 &  $5.58$ \\ 
2 &    $4.27$ \\
3 &   $3.62$
\end{tabular}
\end{center}
\tiny{Parameters:

$a_*^*=b_*^*=c_*^*=1$, $ \oF_1=5,\oF_2=4,\oF_3=3$ \\
 $ \oS_1=8,\oS_2=9,{\mathbf{ \oS_3=10}}$
}
\end{minipage}
\hspace{0.6cm}
\begin{minipage}[!tbp]{0.43\textwidth}
\begin{center}
\includegraphics[scale=0.6]{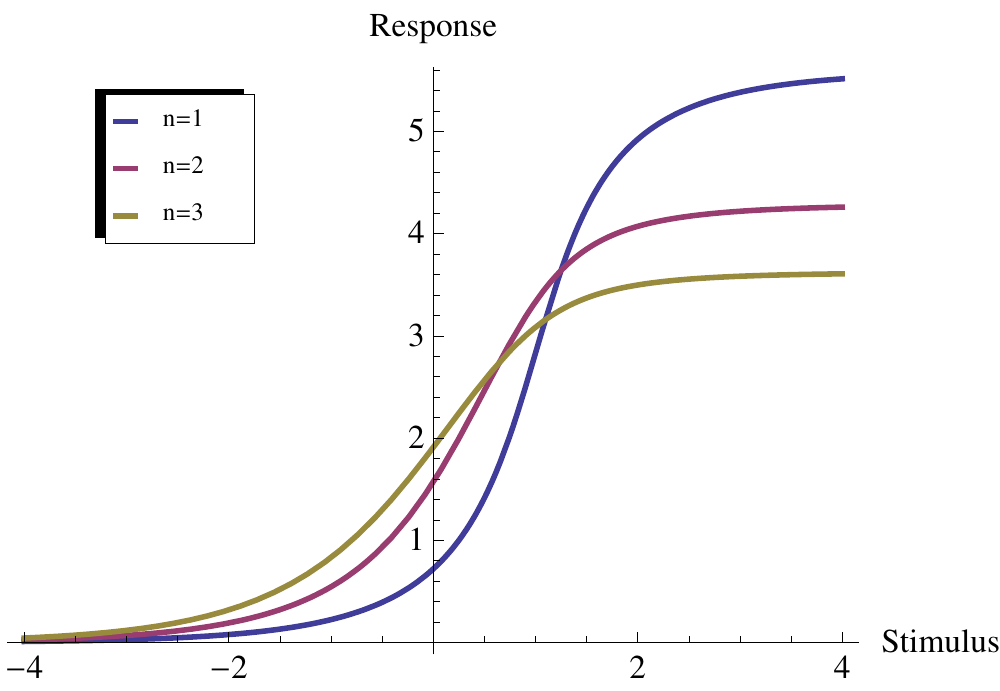}
\tiny{Stimulus-response curves in semi-log scale ($\log(\oE)$ vs $S_n^1$) for different cascade lengths $n=1,2,3$. }
\end{center}
\end{minipage}
}
\caption{Maximal values of the response $S_3^1$ for different cascade lengths. }\label{roc}
\end{figure}

 It follows that each layer of the cascade reduces the maximal response by means of sequestration of substrate in the intermediate complexes. In Figure \ref{roc} this effect is illustrated for a cascade with $n=3$ layers. The maximal response $\sigma_n$ ($=\beta_0$) of the last modified substrate is given in terms of the length of the cascade. In this system all rate constants are the same and the maximal responses are $5.58$, $4.27$, and  $3.62$, respectively, which are much lower than the upper bound set by the total amount (fixed to $10$).

\subsection{Variation due to protein abundance}\label{downup}

Our setting is well suited to study how alterations in  protein  abundance in some layer (e.g.~due to noise, or self-regulation) affects the system. The biological interest in this type of analytical inquiry is discussed in \cite{bluthgen-switch} under the name ``slow regulation''. It  is pointed out that expression of phosphoproteins is altered during many biological processes. 

Variation in signaling protein abundance corresponds to variation in the total amount of substrate in some layer. 
For slow  alterations, the cascade will readjust to a new steady state. Theorems \ref{down} and \ref{alternation} provide a clear picture of how modified and unmodified substrate concentrations  of all layers are affected.

Theorem \ref{down} tells us that an increase in the total amount of substrate in some layer (say, $i$) is transmitted downstream at steady state as an increase in the concentrations of the  modified substrates. Indeed, the effect is equivalent to increasing the stimulus in the smaller cascade consisting of the layers below the one undergoing variation. However,  these layers have fixed total amounts and therefore the modified substrate in layer $j> i$ is still bounded above by $\alpha_j$. 
As the total amount of substrate is increased indefinitely in the layer of modification,  the modified substrate might  also increase indefinitely  or come to a halt at some specific value. In the latter case, the substrate accumulates in unmodified form. Whether one type of outcome or the other occurs, depends on the total amount of  phosphatase in comparison with the total amount of substrate in the previous layer.

\begin{figure}[t]
\framebox[\textwidth]{
\begin{minipage}[h]{0.5\textwidth}
\fcolorbox{black}{white!80!gray}{\textbf{Upstream and downstream variability}}

\vspace{0.4cm}
\centering
\subfloat[
$\oF_3=3$. The substrate $S_3^1$ goes to infinity and this causes $S_4^1$ to approach $\alpha_4=3.27$. ]{\includegraphics[scale=0.5]{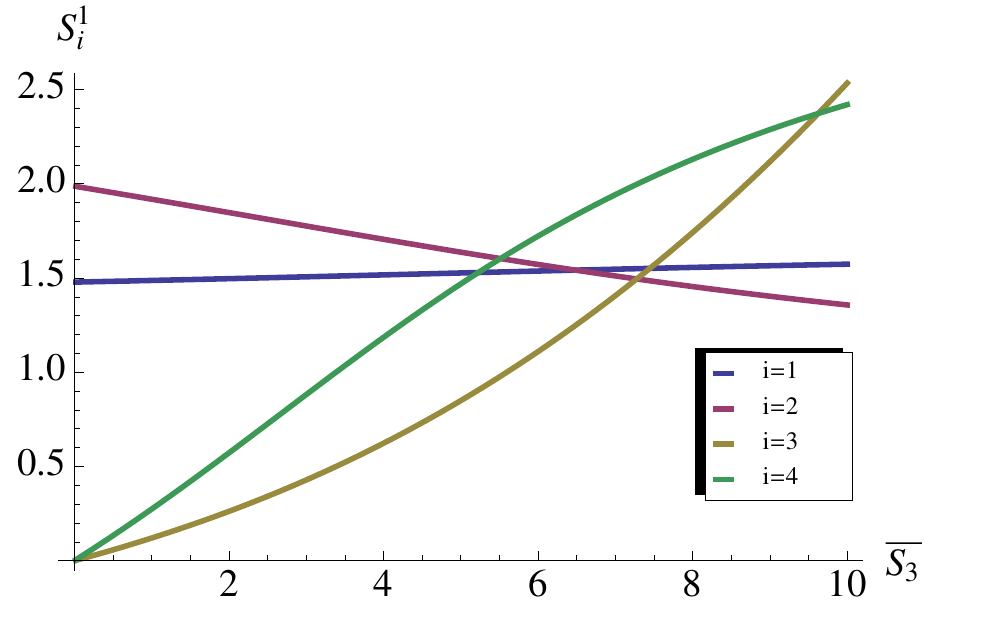}}
\end{minipage}

\begin{minipage}[h]{0.45\textwidth}
\centering
\begin{minipage}[h]{0.83\textwidth}
\tiny{
$a_*^*=b_*^*=c^*_*=1;$ 
$ \oF_*=3,\oE=3,\oS_*=7$ \newline
Increase of $\oS_3$ with different parameter conditions. It causes an increase of $S_1^1,S_3^1$, and $S_4^1$, and a decrease of $S_2^1$.}

\end{minipage}

\centering
\subfloat[
$\oF_3=10$. In this case, $S_3^1$ does not tend to infinity and so $S_4^1$ is kept low too. ]{\includegraphics[scale=0.45]{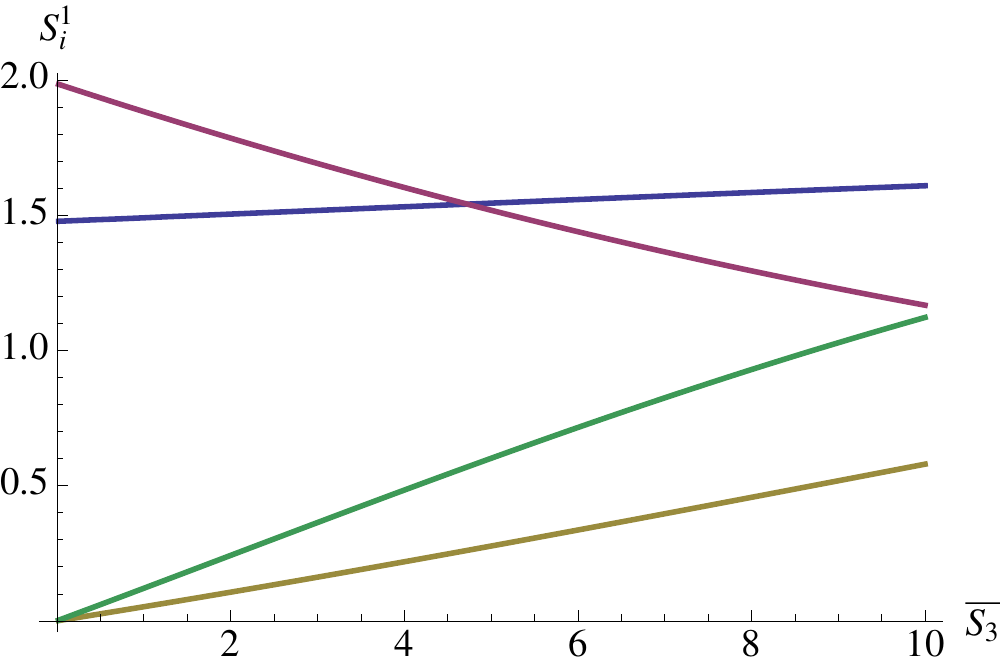}}

\end{minipage}

}
\caption{Illustration of Theorems \ref{down} and \ref{alternation} for $n=4$. 
 }\label{alternate}

\end{figure}

Theorem \ref{alternation} tells us that an increase in the total amount of substrate in layer $i$  is transmitted upstream in an alternating way. In this case,  the concentration of the modified substrate in a  layer $j$ above layer $i$ depends on  the variation in the total amount,  $\oS_j-Y_{j+1}^0$. Therefore, how large the change is, depends on the change in sequestered substrate itself $Y_{j+1}^0$. In particular, this suggests that the effect of  variation is almost negligible in the layers far upstream of the varied one. 
This is illustrated in Figure \ref{alternate}.

Regarding the concentration of unmodified substrate, then the effect is opposite to that of the modified substrate.
By  \eqref{reduced} $S_i^0 =\lambda_i Y_i^0/ (\gamma_i S_{i-1}^1)$, thus an increase in $\oS_i$ causes $Y_i^0$ to increase and $S_{i-1}^1$ to decrease. Hence, $S_i^0$ increases as well. However, for $j\not= i$, the total amount $\oS_j$ is constant and consequently, the concentrations $S_j^0$ change  oppositely to that of $S_j^1$.

\subsection{Stimulus-response curves}\label{hillcoeff}

In our model, variation in the total amount of enzyme (stimulus) is a special case of variation in the total amount of substrate in some layer. Stimulus-response curves have been the focus of many papers, e.g.~\cite{Bluthgen-sequestration,Heinrich-kinase,bluthgen-switch,Vondriska-cascade,Ventura-Hidden}. A recurrent topic  is  (ultra)sensitivity of a system, or the capacity by which a signal transforms the output in a switch-like mode. An analysis of this requires a quantification of the system's switch behavior and  different measures  have been applied here \cite{G-PNAS}. 

Measures of sensitivity rely typically on the amount of enzyme (stimulus) required to achieve 90\% of the maximal steady state value of the modified substrate (response) compared to that required to achieve 10\%.
 Recall that we define $\sigma_n$ to be the maximal  value of the response $S_n^1$. For $M$ in $[0,1]$, let $\oE_M$ be the amount of $\oE$ that corresponds to a steady state value of $S_n^1=M\sigma_n$.  For example,  90\% of the maximal response corresponds to a steady state value of  $0.9\sigma_n$. According to Theorem \ref{stimulus},  the total amount of enzyme required is determined as $\oE_{0.9}=r(0.9\sigma_n)$. 

Therefore, our results provide a way to analyze and compute the sensitive or switch-like character of a signaling cascade. In particular, the \emph{response coefficient} (also called cooperativity index) is given as $R=\oE_{0.9}/\oE_{0.1}$ \cite{Gold-Kosh-81}, the \emph{switch value} as $\oE_{0.9}-\oE_{0.1}$ \cite{G-PNAS} and the \emph{Hill coefficient} as $n_H=\log (81)/\log (\oE_{0.9}/\oE_{0.1})$ \cite{Huang-Ferrell}.

Other measures of sensitivity, probably more suited for an analytical study, escape from our control because the analytical form of the stimulus-response function is required. In our case, the  stimulus is given in terms of the rational function $r$ evaluated in the response (Theorem~\ref{stimulus}), hence the stimulus-response curve is the inverse of $r$. We do not have an expression for the inverse and can only evaluate this through tabulation of points $(M\sigma_n,\oE_M)=(M\sigma_n,r(M\sigma_n))$. One alternative measure is the \emph{control curve}  given as $(x/f)(df/dx)$, where $f$ is the stimulus-response function of the variable $x$ \cite{G-PNAS}. 
For $f=r^{-1}$, a plot of the control curve could be created similarly to that of $r^{-1}$ by tabulation and using that the derivative is one over the derivative of $r$.

\begin{figure}[t]
\framebox[\textwidth]{
\begin{minipage}[h]{0.5\textwidth}
\centering
\begin{minipage}[h]{0.9\textwidth}
\fcolorbox{black}{white!80!gray}{\textbf{Stimulus-response curve}}
\end{minipage}

\medskip
\begin{minipage}[h]{0.9\textwidth}
Shift of the  stimulus-response curves of the different modified substrates $S^1_i$, $i=1,2,3$, for $n=3$. Substrate values are normalized to allow comparison. Graphic in semi-log scale.

\bigskip

\tiny{Parameters: \\
$a_*^*=b_*^*=c_*^*=1$, $ \oF_1=5,\oF_2=4,\oF_3=3$ \\
 $ \oS_1=8,\oS_2=9, \oS_3=10$
}

\vspace{0.3cm}

\end{minipage}
\end{minipage}
\hspace{0.5cm}
\begin{minipage}[h]{0.45\textwidth}\centering
{\includegraphics[scale=0.6]{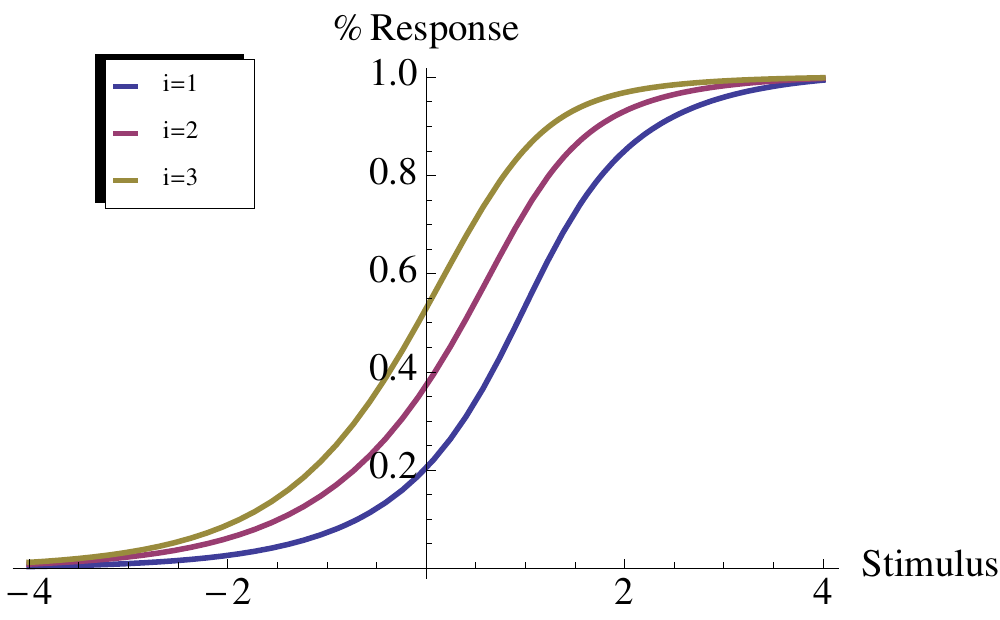}}
\end{minipage}

}
\caption{Stimulus-response curves in semi-log scale.  }\label{stimulusgraphics}

\end{figure}

Further, we have studied the shift in  stimulus-response curves for  modified substrates in different layers, as suggested in \cite[Fig. 6]{Ventura-Hidden}. For that, we note that  $\sigma_i=f_i(\sigma_n)$ is the maximal value of the modified substrate in the $i$-th layer (Proposition~\ref{rationalfunctions} and Theorem \ref{stimulus}). Then, if $M\in [0,1]$, we consider $\oE_M^i$ to be the amount of $\oE$ that corresponds to a steady state value of $S_i^1=M\sigma_i$, that is, $M$ times the maximal response. We   have that $\oE_M^i = r(M f_i(\sigma_n))$ and  for any set of parameters 
\begin{equation}\label{shift}
\oE^{i+1}_M<\oE_M^i
\end{equation}
(proved in  Appendix).
That is, the stimulus-response curves for the different modified substrates are shifted from right to left with increasing  index of the layer. Consequently, in order to achieve maximal response, $S_n^1$ requires the least kinase while $S_1^1$ requires the most. This is illustrated in Figure \ref{stimulusgraphics}.

\subsection{Enzyme competition} \label{enzcomp}

The one-site modification cycle considered here is driven by two opposing processes, modification and demodification, catalyzed by a kinase and a phosphatase, respectively. Both enzymes compete for the substrate and one expects that if the kinase is in excess over the phosphatase, then the system shifts in favor of the modified substrate. The results in  section $\S$\ref{math_sec} allow for a precise statement of this fact, as well as its extension to the entire cascade. 

It follows from  Proposition \ref{enzymebehavior}, that the steady state value of the modified substrate is an increasing function of the total amount of substrate. The modified substrate takes values in  
 $\overline{\R}_{+}$  for $\Delta\leq 0$, and $[0,\oE/\Delta)$ for $\Delta>0$, where  $\Delta = \gamma\oF-\delta\oE=\delta((c^1/c^0) \oF - \oE)$.
Hence $\Delta \leq 0$ only if $T=c^1 \oF - c^0 \oE\leq 0$.
This implies that as the total amount of substrate is increased, the steady state concentration of the modified substrate  increases indefinitely or approaches the asymptotic value $\oE/\Delta$, depending on the  initial amount of enzymes  and the dissociation constants. Additionally,  from \eqref{Y0rel} and the conservation law for $\oE$, we have that $S^0=\lambda\overline{F}S^1/(\overline{E}-\Delta S^1)$, which is an increasing function of $S^1$. We conclude that when $\oS$ increases, $S^0$  tends to $-\lambda \overline{F} / \Delta>0$ if $T< 0$ and to $+\infty$ otherwise. Only when $T=0$, both the unmodified and modified substrate concentrations become large at the same time.

For cascades of length $n>1$, the same competition is observed when the total substrate concentrations $\oS_1,\dots,\oS_{n-1}$ are fixed and $\oS_n$ is increased. 
Theorem \ref{onesteadystate} shows that  $Y_n^0$ tends to 
$\min((c^1_n/c_n^0)\oF_n,\oS_{n-1})$ as $\oS_n$ becomes large. In addition,  $S_{n-1}^1$ is a function of $Y_n^0$, $f(Y_n^0)$ with $f(\oS_{n-1})=0$, and hence by \eqref{reduced} and \eqref{Y0rel},
$S_n^0=(\lambda_n Y_n^0)/(\gamma_n f(Y^0_n))$ and $ S^{1}_{n}= Y_n^0/( \gamma_n \overline{F}_{n} - \delta_n Y_n^0)$.Therefore, the sign of $T_n=c^1_n\oF_n - c^0_n \oS_{n-1}$ determines the limit value of $Y_n^0$, and thus the asymptotic values of the modified and unmodified substrates. 
Indeed, if $T<0$ then $S_n^1$ tends to infinity, while $S_n^0$ tends to the asymptotic value; when $T>0$ the opposite behavior is observed. Only when $T=0$ both substrates become large at the same time.

\section{Concluding remarks}\label{concluding}

In this paper we have studied   signaling cascades consisting of $n$ identical layers of modification. We have
shown that it is possible to draw biological relevant conclusions from  analytical aspects of rational functions describing the system.
The complexity of the system is reduced by splitting the cascade into smaller cascades. 
 
In particular,  we have shown that signaling cascades  with one PTM in each layer cannot exhibit multistability for any rate constants or total initial amounts of substrates and enzymes.  It is well known that more complex signaling cascades, like the MAPK cascade, can exhibit bistability, e.g. \cite{Ferrell-bistability,Markevich-mapk,Qiao-bistability}.  Since multistability in PTM only arises when there are multiple modification sites, we conclude that multistability in a signaling cascade (without feedback) must be a consequence of the presence of multiple steady states in some of its layers, and cannot  be an effect of the cascade itself. 

Further, we have shown that stimulus-response curves can be obtained as  inverse functions of certain rational functions which are explicitly given. Stimulus-response curves are important as they provide  theoretical interpretation of the behavior of the system. Also,  points on  a stimulus-response curve can be determined experimentally and  used to draw inference on the kinetic parameters of the system, e.g. \cite{Meinke-zero}. 
However, the stimulus-response curve cannot be given in the form of a closed analytical expression. To find the inverse of the corresponding rational function we need to determine the roots of a high-degree polynomial. For $n=3$, the polynomial describing the steady states of the system  has already degree $17$ rendering exact analysis difficult.

Finally, we have studied sequestration. Variation in the levels of the total amounts of substrates influence the steady state of the system. Downstream, variation is transmitted positively, while upstream, variation is transmitted in an alternating fashion. As a consequence, 
the modified substrate in the last layer is always positively influenced by variations in any layer.

The framework presented here is quite flexible and allows us to address questions about the system theoretically without restoring to simulation. The way we use the modularity of the cascade to derive properties of the cascade and to determine the number of steady states suggests that our results could be extended to signaling cascades with layers consisting of other reaction systems, as long as we have some mathematical knowledge about the variation of the species with respect to the total amounts. Such a study should aim to relate the number of steady states of the cascade to the number of steady states arising from each  isolated layer. As a consequence, a better understanding of the emergence of multistability in signaling cascades may be provided.  In particular, we believe that the ideas developed in the paper could be useful for studying more complex cascades, like the MAPK cascade, and for example help to better elucidate the parameter regions of bistability \cite{conradi-mapk,Ortega-bistability}.

\appendix

\section{Proofs}

\begin{proof}[Lemma \ref{lemma_pos}]
We use  \eqref{totalamounts} and \eqref{reduced} repeatedly without further reference.
(i) If $E=0$, then  $Y_1^0=0$ (since $E=S_0^1$ by convention) and hence $\oE=E+Y_1^0=0$. (ii) If $F_i=0$, then $\oF_i=0$. If $1+\delta_iS_i^1=0$, then  $Y_i^1=-F_i$ and hence $\oF_i=0$. (iii) If $S_i^1=0$, then $Y_i^0=Y_i^1=0$. Also either  (a) $S_i^0=0$ or (b) $S_{i-1}^1=0$. If $S_i^0=0$ (a), then $\oS_i=S_i^0+S_i^1+Y_i^0+Y_i^1+Y_{i+1}^0=Y_{i+1}^0$. For $i=n$, $Y_{n+1}^0=0$, while for $i<n$, $Y_{i+1}^0=(\gamma_{i+1}/\lambda_{i+1})S_i^1S_{i+1}^0=0$; hence in both cases $\oS_i=0$. If  $S_{i-1}^1=0$ (b), then repeat the argument until either $S_0^1=E=0$ (hence $\oE=0$, according to (i)\,) or  $S_j^0=0$ for some $j<i$. In the latter case it follows that $\oS_j=0$, using (a). (iv) If $S_i^0=0$, then $Y_i^0=0$, and either $\oF_i=0$ or $S_i^1=0$. The result now follows from (iii) (a).
\qed\end{proof}

\begin{proof}[Proposition \ref{enzymebehavior}] 

The function $\varphi$ is well defined for $S^1=0$ with value $\oS=0$, and it is continuous outside the roots of the denominator. The only possibly positive root of the denominator is $\eta=\oE/\Delta$, provided $\Delta\neq 0$. 

The derivative of $\varphi$ with respect to $S^1$ is  $\varphi'= 1+ N/[ q_{1}(S^1)^2q_{2}(S^1)^2]$, where
$N = \lambda\overline{F} q_{1}^2 (q_{2}-q_{2}'S^1) + \epsilon q_{2}^2 (q_{1}-q_{1}'S^1) $ and
 $\epsilon = \gamma+\delta \overline{F}$. Since $q_{2}-q_{2}'S^1=\overline{E}$ and $q_{1}-q_{1}'S^1=1$, 
then $N>0$. Therefore, $\varphi'>0$ for all values of $S^1$ for which $\varphi$ is defined, and hence it is an increasing function.

In the following we rely on \eqref{Y0rel} and Lemma \ref{lemma_pos}. Since $\oE> 0$,  we have $E\not=0$ at steady state. Therefore  in order for $E>0$, we need
\begin{equation}\label{E0}
E=\oE-Y^0=\overline{E} - \frac{ \gamma  \overline{F} S^1}{1+\delta S^1} >0 \Leftrightarrow \overline{E} > \Delta S^1. 
\end{equation}
For values of $S^1$ for which \eqref{E0} holds, we have $q_{2}(S^1)>0$ and  $q_{1}(S^1)>0$, hence $p(S^1)> 0$, and $\varphi(S^1)=0$  if and only if $S^1=0$. Therefore, $\oS=\varphi(S^1)>0$ if \eqref{E0} holds. Define the remaining quantities ($S^0,Y^0,Y^1,F$) through  \eqref{Y0rel}. These quantities are all positive.
Therefore, a positive value of $S^1$ provides a steady state in $D$ if and only if  \eqref{bS} and \eqref{E0} are satisfied. The question is whether for every value of $\overline{S}$ there is a unique value of $S^1$ satisfying equations \eqref{bS} and \eqref{E0}. 

 Let $\overline{\varphi}$ be the restriction of $\varphi$ to the set $\Gamma$ of values  $S^1\geq 0$ satisfying  \eqref{E0}. 
There are two possible scenarios: 
\begin{enumerate}[(i)]
\item If $\Delta\leq 0$, then   \eqref{E0} is satisfied for all values of $S^1$ and so $\Gamma=\overline{\R}_{+}$. In this case, there is no positive root of $q_2(S^1)$ and hence  $\overline{\varphi}$ is a continuous increasing function in $\Gamma$.  
\item If $\Delta>0$, then  \eqref{E0} is satisfied if $0\leq S^1 < \eta=\oE/\Delta$. Since  the only positive root of the denominator of $\varphi$ is $ \eta$,  we have $\overline{\varphi}$ is a  continuous increasing function in $\Gamma=[0,\eta)$. 
\end{enumerate}
In both cases, the image of $\overline{\varphi}$ is the set of positive real numbers $\overline{\R}_{+}$. By Lemma \ref{rational}, the inverse function, $\psi$, is also a  continuous increasing function of $\overline{S}$ in $\overline{\R}_{+}$ with image set $\Gamma$.  It follows that there is a unique steady state of the system in $D$ for positive  total amounts $\overline{E},\overline{F}$, and  $\overline{S}$.
\qed\end{proof}

\begin{proof}[Proposition \ref{functiong}]
It is straightforward to check that the derivative of $g=p_1/p_2$ as a function of $Y^{0}$ is always positive and hence $g$ is increasing, whenever it is defined. Recall that $
p_1(y)  =  \lambda y(\xi- y)$, $p_2(y) = (\delta+\gamma)y^2 - \gamma (1/\delta +\oF+\xi + \oS)y + \gamma\xi\oS,$  with $\xi=\gamma\oF/\delta$.
The signs of the  coefficients of $p_2(y)$ indicate that $p_2(y)$ has either two positive roots or two non-real conjugate complex roots. For $y=0$, $p_2(0)>0$, while for $y=\xi$, $p_2(\xi)=-\gamma\xi/\delta<0$. Therefore, there is at least one positive root $\alpha_1\in(0,\xi)$.
When $y$ tends to infinity, $p_2(y)$ becomes positive, and thus the other real root $\alpha_2$ satisfies $\alpha_2>\xi>\alpha_1$. It follows that  $p_2(y)$ is positive in $[0,\alpha_1)$ and $(\alpha_2,+\infty)$ and negative in $(\alpha_1,\alpha_2)$.
Note that $p_1(y)$ has constant positive sign in $(0,\xi)$ and takes  negative values in $(\xi,+\infty)$. Hence, $E$ is positive only if $Y^0\in (0,\alpha_1)\cup (\xi,\alpha_2)$.

If $Y^0> \xi$, then $Y^1> \oF$ meaning that $F<0$. Therefore, any steady state  in $D$  must satisfy $Y^0\in (0,\xi)$.  It is left to the reader to check that  the derived steady state values for $Y^1,S^1,S^0,F$ are in $D$ for  $Y^0\in (0,\alpha_1)$.

In the interval $[0,\alpha_1)$, $g$ is a positive continuous increasing function of $Y^{0}$, which is zero when $Y^{0}=0$ and tends to infinity when the root $\alpha_1$ is approached. 
\qed

\end{proof}

\begin{proof}[Theorem \ref{onesteadystate}]
We are going to prove the theorem by induction on $n$. For $n=1$, (ii) follows directly from Proposition \ref{enzymebehavior}; (i) follows by composition of $\psi_1=\psi$ in Proposition \ref{enzymebehavior} with the expression for $Y_1^0$ in  \eqref{Y0rel}. Next, assume that the theorem holds for $n-1$ for some $n>1$.

Consider the splitting of the cascade of length $n$ into the last layer $C_{n}$  and the first $n-1$ layers $C$. 
By induction hypothesis, for every value of $Y_{n}^{0}\in [0,\oS_{n-1})$,  there is exactly one steady state of $C$.
In addition,  $S^{1}_{n-1}$ is given by a  continuous increasing function $\psi_{n-1}$ of the total amount of substrate in the last layer. Hence $S_{n-1}^1=f(Y_n^0)$ with $f$  a decreasing function of $Y^{0}_n$ in the interval $[0,\oS_{n-1})$. Further, $f$ can be extended continuously to $\oS_{n-1}$ with $f(\oS_{n-1})=0$, since $\psi_{n-1}(0)=0$. Note that $f$ is independent of $\oS_n$.

On the other hand, we noticed in the main text that $S^{1}_{n-1}$ is given by a  continuous increasing function $g_n$ of $Y^{0}_n$, for $Y_n^0$ in the valid interval $[0,\alpha_1)$. The function  $g_n$  is zero when $Y^{0}_n=0$ and tends to infinity when the root $\alpha_1$ is approached. 

Therefore, there are two continuous functions of $Y^{0}_n$, $f(Y^0_n)$ and $g(Y^0_n)$, describing the steady state value of $S^{1}_{n-1}$. Both of them take positive values, but  $f(Y^0_n)$ is decreasing, while $g(Y^0_n)$ is increasing in $[0,\alpha_1)$. In addition, $f(\oS_{n-1})=0$ and $f(0)>0$. Since the image set of $g$ on $[0,\alpha_1)$ is $\overline{\R}_{+}$, there exists a unique value of $Y^0_n$ for which $f(Y^0_n)=g(Y^0_n)$, and therefore there is a unique BMSS.
Note that the intersection point satisfies $0<Y_n^0<\min(\xi_n,\oS_{n-1})$, $\xi_n=\gamma_n \oF_n/\delta_n$.

What remains to  prove is that  the statements (i) and (ii) also are  satisfied for $n$. 
(i)  The steady state is  the unique value for which $f(y)=p_1(y)/p_2(y)$. Note that $p_2(y)$ is  a linear polynomial in $\overline{S}=\overline{S}_{n}$ with coefficients in $\R[y]$. Therefore,  $p_2(y)=a(y) + b(y)\overline{S}$ with $b(y)=\gamma_n (\gamma_n \oF_n - \delta_n y)/\delta_n>0$. The steady state solution is then given by
\begin{equation*}
\overline{S} = \varphi_n(y):=\frac{p_1(y)-a(y)f(y)}{b(y)f(y)}.
\end{equation*}
This function is continuous in the interval $[0,\mu_n)$, where $\mu_n=\min(\xi_n,\oS_{n-1})$.
In addition, its derivative is positive. Indeed, 
$$\varphi'_n = \frac{1}{b^2f^2} \big(-p_1bf' + f^2(ab'-a'b) + f (p_1'b-p_1b')\big).$$
We have that $f'<0$ and $p_1,b>0$. Also,  $ab'-a'b=\oF \gamma^3/\delta^2 +\gamma(\delta+\gamma)(\oF \gamma/\delta-  y)^2>0$
(suppressing the subindices) and $p_1'b-p_1b'=\gamma\lambda(\xi-y)^2>0$. Consequently, $\varphi'_n>0$ and $\varphi_n$ is increasing.
Therefore, by Lemma \ref{rational}, the intersection point is given by a  continuous increasing function of $\overline{S}$. It proves (i). (ii) From \eqref{Y0rel},
\begin{equation*}
S^{1}_{n}=\chi(Y_n^0)=\frac{ Y_n^0 }{ \gamma_n  \overline{F}_{n} - \delta_n Y_n^0},
\end{equation*}
which is an increasing function of $Y_n^0$; hence by composition of $\chi$ with $\varphi^{-1}_n$, $S_n^1$ is an increasing function of $\oS_n$. The inverse of this function is $\varphi_n(\gamma_n \oF_nS_n^1/(1+\delta_nS_n^1))$,
which is a rational function of $S_n^1$, as desired. It is defined on $\Gamma_n=[0,\chi(\mu_n))$, which is $+\infty$ if $\mu_n=\xi_n$, and $\oS_{n-1}/\Delta_n$ otherwise. 
The proof of Theorem \ref{onesteadystate} is completed.
\qed\end{proof}

\begin{proof}[Proposition \ref{rationalfunctions}]
The proof is by decreasing induction starting from $i=n-1$. The discussion above Proposition \ref{rationalfunctions} shows that there is a rational function $S_{n-1}^1=f_{n-1}(S_n^1)$ satisfying (i) with $\beta_{n-1}=\alpha_n$. To prove that the image is $\overline{\R}_{+}$, note  that  the derivative of $f_{n-1}$ with respect to $S_n^1$ is always positive, and therefore $S_{n-1}^1$ is a continuous increasing function of $S_n^1$ (for $\oS_n,\oF_n$ fixed), which tends to infinity as $S_n^1$ tends to $\alpha_n$, and vanishes if  $S_n^1=0$. (Note that $f_{n-1}$ only depends on $\oF_n$ and $\oS_n$, so when $S_n^1$ (at steady state)
 tends to $\alpha_n$, it implicitely implies that $\oS_{n-1}$ tends to infinity.) (ii) follows from the expression of $d_n(x,0)$, and (iii) is likewise fulfilled. 

Now assume the proposition is true for all $j$ satisfying $i\leq j \leq n-1$ for some $i$. Consider $i-1$.
By induction hypothesis,  $S_j^1=f_j(S_n^1)$ are increasing rational functions for all $j\geq i$, 
and so, by \eqref{Y0rel}, $Y_{j+1}^0=f^Y_{j+1}(S_n^1)$ is an increasing rational function of $S_n^1$.
Hence by \eqref{rationalsi}, $S_{i-1}^1$ is a rational function of $S_n^1$ given by $f_{i-1}(S_n^1)=g_i\big(f_i(S_n^1),f^Y_{i+1}(S_n^1)\big)$.
If $i+1=n$, then $f_n$ is the identity function. 
Note that $f_{n-1}$ depends on $\oS_n$ and $\oF_n$, but not on any other  total amounts. Similarly, $g_i$ depends on $\oS_i$ and $\oF_i$. By induction $f_{i-1}$ only depends on $\oS_j$ and $\oF_j$, $j\geq i$. 

By induction hypothesis, $S_{i}^1$, $S_{i+1}^1$ and $Y_{i+1}^0$ are   continuous increasing functions of $S_n^1$ defined on the interval $[0,\beta_i)$. In addition, as $S_n^1$ tends to $\beta_i$, $S_i^1$ tends to infinity, $S_{i+1}^1$ tends to the value $\eta=f_{i+1}(\beta_i)<+\infty$ (because $\beta_i<\beta_{i+1}$) 
and so $Y_{i+1}^0$   tends to  $\eta^Y=f_{i+1}^Y(\beta_i)<+\infty$.

The function $\rho_{i}(Y_{i+1}^0)$ is a continuous decreasing  function of $Y_{i+1}^0$, ranging from $\alpha_i$ to $0$, defined as the positive root of the  polynomial $d_i(x,Y_{i+1}^0)$.
Let $\overline{\rho}_i=\rho_i\circ f_{i+1}^Y$.
Since $Y_{i+1}^0$ is increasing as function of $S_{n}^1$, then $\overline{\rho}_i(S_n^1)$ is decreasing, because $\rho_{i}(Y_{i+1}^0)$ is. 
In order for $S_{i-1}^1$ to be well defined, it must be that $S_i^1=f_i(S_n^1)< \overline{\rho}_i(S_n^1)$ (it is negative otherwise). Therefore, the possible steady state values of $S_n^1$ must satisfy $f_i(S_n^1)-\overline{\rho}_i(S_n^1)<0$. The function $f_i-\overline{\rho}_i$ is a continuous increasing function that takes the value $-\alpha_i$ at $S_n^1=0$ ($Y_{i+1}^0=0$).
On the other hand, when $S_n^1$ tends to $\beta_i$, the function $f_i-\overline{\rho}_i$ tends to $+\infty$. 
 Therefore, there exists a unique value of $S_n^1$, $\beta_{i-1}<\beta_i$,   for which $f_i-\overline{\rho}_i=0$. This value is the upper limit for $S_n^1$. Thus, a necessary condition for valid steady states is that $S_n^1\in [0,\beta_{i-1})$. Note that  $f_i(\beta_{i-1})=\overline{\rho}_i(\beta_{i-1})$ is the $S_i^1$-value corresponding to the first positive root of  $d_{i}(S_n^1,f_{i+1}^Y(S_n^1))$.
 
 It follows from \eqref{rationalsi} that $S_{i-1}^1$ is an increasing rational function of $S_n^1\in [0,\beta_{i-1})$, because the denominator is non-zero for values $S_n^1<\beta_{i-1}$. It proves the first part of the statement.  Since for positive values of $S_n^1$ smaller than $\beta_{i-1}$ the denominator is non-zero, we conclude that $\beta_{i-1}$ corresponds to the first positive zero of the denominator. In addition, when $S_n^1$ tends to $\beta_{i-1}$, the denominator of \eqref{rationalsi} tends to zero, while the numerator tends to some positive real number (since $\beta_{i-1}<\beta_i$) and hence $S_{i-1}^1$  tends to infinity showing (i).
 
To see (ii), note that $f_i$ does not depend on $\oS_i$, while $\overline{\rho}_i$ does. If $\alpha$ is a fixed value of $S_n^1$, and $\oS_i<\oS_i'$, then we have $\overline{\rho}_i(\alpha,\oS_i)<\overline{\rho}_i(\alpha,\oS_i')$. Therefore, the intersection point of the two curves increases if $\oS_i$ does. In addition, $\overline{\rho}_i$ tends to infinity as $\oS_i$ does (consider the expression for the positive root), and therefore intersection points as close to $\beta_{i}$ as desired can be obtained.
  \qed \end{proof}
 
 \begin{proof}[Theorem \ref{down}]
 From \eqref{Y0rel} and Proposition \ref{rationalfunctions}, we have that $S_i^1$ and $Y_{i+1}^0$ are  continuous increasing functions of $S_n^1$. Hence,  by the inverse function theorem, $S_i^1$ is a continuous increasing function of  $Y_{i+1}^0$. Call this function  $\Psi_i$. It does not depend on  $\oS_{i}$. 

On the other hand, the proof of Theorem \ref{onesteadystate} demonstrates 
that for every fixed total amount $\oS_i$, $S_i^1$ is a continuous decreasing function of $Y_{i+1}^0$, $\Phi_i(\oS_i,Y_{i+1}^0)$. It follows from the fact that the total amount in the $i$-th layer, $\oS_i-Y_{i+1}^0$, decreases with $Y_{i+1}^0$. Therefore, the steady state value of $S_i^1$ is the intersection of these two curves, $\Psi_i(-),\Phi_i(\oS_i,-)$. If now $\oS_i$ is increased, the first curve does not change, while the second one does. For every fixed value $\upsilon$ of $Y_{i+1}^0$, we have that $\Phi_i(\oS_i,\upsilon)<\Phi_i(\oS_i',\upsilon)$ if $\oS_i<\oS_i'$. Therefore,  the steady state value of $S_i^1$ increases as $\oS_i$ does, because we intersect with a growing function. 

The quantities  $S_j^1$, $j\geq i$, are continuous increasing rational functions of $S_n^1$. Further, for $j>i$, the rational functions do not involve $\oS_i$. Hence, $S_n^1,\dots,S_{i+1}^1$ will  also be growing with $\oS_i$. It follows from \eqref{Y0rel} that $Y_j^0$ and $Y_j^1$  also increase, while $S_j^0$ decreases.
\qed \end{proof}

 \begin{proof}[Theorem \ref{alternation}]
 We proved that an increase of $\oS_i$ causes $S_i^1$ to increase and $S_{i-1}^1$ to decrease. Using an induction argument, assume that the statement is true for layers with indices  smaller than some value $k$. Then, if $k=i-2j$, by induction hypothesis $S_{k-1}^1$ decreases if $\oS_i$ increases. Therefore, by \eqref{Y0rel}, $Y_{k-1}^0$ decreases and hence $\oS_k-Y_{k-1}^0$ increases. It follows that $S_{k}^1$ increases, as desired. If $k=i-2j-1$, the argument is analogous.
 \qed \end{proof}

\begin{proof}[Inequality \eqref{shift}] 
Since $S_n^1$ is a continuous increasing function of $S_i^1$, we have that $\oE=r^i(S_i^1)$, where $r^i=r\circ f^{-1}_i$, is given by a continuous increasing function. Additionally, $S_i^1$ is a continuous increasing function of $S_{i+1}^1$ given by $\overline{g}_{i+1}=f_i\circ f_{i+1}^{-1}$ on $[0,\sigma_{i+1})$. We obtain $r^{i+1}=r^{i}\circ \overline{g}_{i+1}$.  Let $\oE_M^i=r^i(M \sigma_i)$ be the value of $\oE$ corresponding to $M\sigma_i$, the maximal value of $S_i^1$. Because  $\sigma_i=\overline{g}_{i+1}(\sigma_{i+1})$, we have $\oE_M^i= r^{i}(M\overline{g}_{i+1}(\sigma_{i+1}))$.  On the other hand,  $\oE_M^{i+1}=r^{i+1}(M\sigma_{i+1})= r^{i}(\overline{g}_{i+1}(M \sigma_{i+1}))$. 

Because $r^{i}$ is an increasing function,  the proposition  follows if  $\overline{g}_{i+1}(M \sigma_{i+1})<M\overline{g}_{i+1}(\sigma_{i+1})$. Let $Y<Y'$ be the  values of $Y_{i+2}^0$ corresponding to $M\sigma_{i+1}$ and $\sigma_{i+1}$, respectively (zero if $i=n-1$). Since $d_{i+1}(M\sigma_{i+1},Y)>d_{i+1}(\sigma_{i+1},Y')$, then by \eqref{rationalsi}, 
$$\overline{g}_{i+1}(M\sigma_{i+1})  = \frac{ \lambda_{i+1}\oF_{i+1} M\sigma_{i+1}}{d_{i+1}(M\sigma_{i+1},Y)}
< M \frac{ \lambda_{i+1}\oF_{i+1} \sigma_{i+1}}{d_{i+1}(\sigma_{i+1},Y')} = M \overline{g}_{i+1}(\sigma_{i+1}),$$
and the inequality is proved.
\qed
\end{proof}

%

\end{document}